\documentclass{aa}  

\usepackage{graphicx,amsmath,multirow}
\usepackage{txfonts}
\bibpunct{(}{)}{;}{a}{}{,} 

\begin{document}

   \title{Single-mode waveguides for GRAVITY}

   \subtitle{II. Single-mode fibers and Fiber Control Unit}

   \author{
        G. Perrin \inst{1}
        \and 
        L. Jocou \inst{2}
        \and
        K. Perraut \inst{2}
        \and
        J.-Ph. Berger \inst{2}
        \and
        R. Dembet \inst{1}
        \and
        P. F\'edou \inst{1}
        \and
        S. Lacour \inst{1}
        \and
        F. Chapron \inst{1}
        \and
        C. Collin \inst{1}
        \and
        S. Poulain \inst{3}
        \and
        V. Cardin \inst{3}
        \and
        F. Joulain \inst{3}
        \and
        F. Eisenhauer \inst{4}
        \and
        X. Haubois \inst{5} 
        \and
        S. Gillessen \inst{4}
        \and
        M. Haug \inst{5}
        \and
        F. Hausmann \inst{4}
        \and
        P. Kervella \inst{1}
        \and
        P. L\'ena \inst{1}
        \and
        M. Lippa \inst{4}
        \and
        O. Pfuhl \inst{5}
        \and
        S. Rabien \inst{4}
        \and
         A. Amorim \inst{6}
        \and
        W. Brandner \inst{7}
        \and
        C. Straubmeier \inst{8}
          }

   \institute{LESIA, Observatoire de Paris, Universit\'e PSL, CNRS, Sorbonne Universit\'e, Universit\'e de Paris, 5 place Jules Janssen, 92195 Meudon, France
        \email{guy.perrin@obspm.fr}
        \and
        Univ. Grenoble Alpes, CNRS, IPAG, 38000 Grenoble, France
        \and
        Le Verre Fluor\'e, rue Gabriel Voisin, 35170 Bruz
        \and
        Max Planck Institute for Extraterrestrial Physics, Giessenbachstrasse, 85741 Garching bei M\"{u}nchen, Germany
        \and 
        European Southern Observatory, Karl-Schwarzschild-Str. 2, 85748 Garching, Germany
        \and
        CENTRA, Instituto Superior Tecnico, Av. Rovisco Pais, 1049-001 Lisboa, Portugal
        \and
        Max Planck Institute for Astronomy, K\"{o}nigstuhl 17, 69117 Heidelberg,Germany
        \and
        I. Physikalisches Institut, Universit\"{a}t zu K\"{o}ln, Z\"{u}lpicher Str. 77, 50937 K\"{o}ln, Germany
             }

   \date{Received June dd, 2023; accepted Mmmm dd, 2023}

 
  \abstract{The second generation Very Large Telescope Interferometer (VLTI) instrument GRAVITY is a two-field infrared interferometer operating in the K band between 1.97 and $2.43\,\mu$m with either the four 8\,m or the four 1.8\,m telescopes of the Very Large Telescope (VLT). Beams collected by the telescopes are corrected with adaptive optics systems and the fringes are stabilized with a fringe-tracking system. A metrology system allows the measurement of internal path lengths in order to achieve high-accuracy astrometry. High sensitivity and high interferometric accuracy are achieved thanks to (i) correction of the turbulent phase, (ii) the use of low-noise detectors, and (iii) the optimization of photometric and coherence throughput. Beam combination and most of the beam transport are performed with single-mode waveguides in vacuum and at low temperature. In this paper, we present the functions and performance achieved with weakly birefringent standard single-mode fiber systems in GRAVITY. Fibered differential delay lines (FDDLs) are used to dynamically compensate for up to 6\,mm of delay between the science and reference targets. Fibered polarization rotators allow us to align polarizations in the instrument and make the single-mode beam combiner close to polarization neutral. The single-mode fiber system exhibits very low birefringence { (less than $23^{\circ}$)}, very low attenuation { (3.6-7\,dB/km across the K band),} and optimized differential dispersion { (less than $2.04\,\mu \mathrm{rad \, cm}^2$ at zero extension of the FDDLs)}. As a consequence, the typical fringe contrast losses due to the single-mode fibers are 6\%\ to 10\% in the lowest-resolution mode and 5\% in the medium- and high-resolution modes of the instrument for a photometric throughput{ of the fiber chain} of the order of 90\%. { There is no equivalent of this fiber system to route and modally filter beams with delay and polarization control in any other K-band beamcombiner}.}
 


   \maketitle
%



   \begin{figure*}[t]
   \centering
   \includegraphics{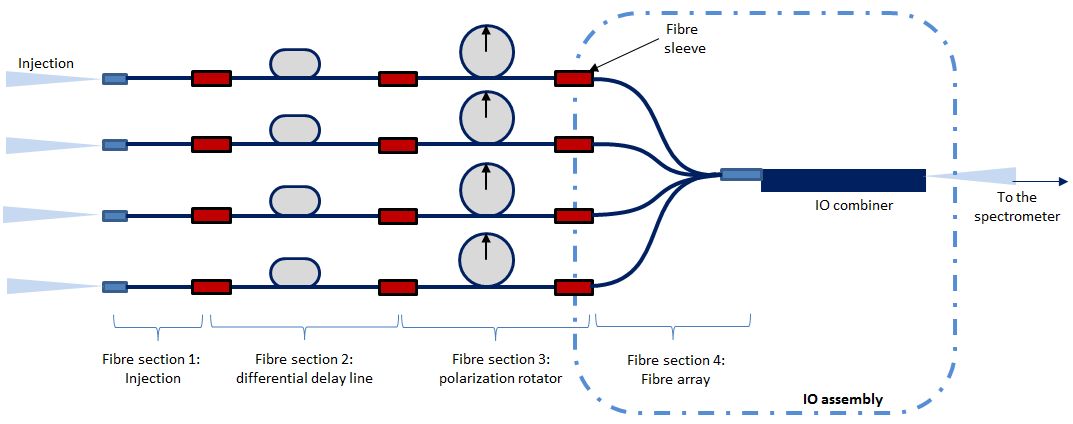}
   \caption{Overview of the GRAVITY chain of single-mode waveguides. The light coming from the telescopes is injected in the Fiber Coupler fibers inside the cryostat of GRAVTY (first section). Each fiber is then connected to a FDDL (section 2) and to a Fibered Polarization Rotator (section 3). The light is then fed to the V-groove fibers connected to the integrated optics beam combiner (section 4 inside the dash-dotted line box). All work at cold temperatures (240 K for the first three sections and 200\,K for the integrated optics beam combiner part). This figure was first presented in \citet{Perraut2018}.}
              \label{Fig:chain}%
    \end{figure*}

\section{Introduction}
\label{sec:introduction}
\nolinenumbers
GRAVITY is a second-generation instrument of the Very Large Telescope Interferometer (VLTI) of the European Southern Observatory (ESO). It is operated in the K band between 2 and $2.4\,\mu$m and allows long-baseline interferometry with either the four 8\,m Unit Telescopes (UTs) or the four 1.8\,m Auxiliary Telescopes (ATs). Adaptive-optics systems correct for turbulence in the individual pupils and fringes are stabilized with a fringe-tracking system. When not the science target itself, a reference source for fringe tracking can be selected in a small field around the science target (2 arcsec with {the} UTs, 6 arcsec with the ATs) in order to perform double-field interferometry. A metrology system allows the two fringe systems to be connected in order to perform high-precision narrow-angle astrometry or phase-referenced imaging. Images can also be reconstructed with GRAVITY using closure phases. The instrument is presented in \citet{GRAVITY_1st_light_2017}. The present paper outlines the rationale for using single-mode fibers in GRAVITY and the characteristics of the fibers used in the instrument as well as of the functions performed by the fibers; it is the second in a series of papers on the single-mode waveguides of GRAVITY and follows \citet{Perraut2018}.

{ The use of single-mode fibers for astronomical interferometry was first suggested by \citet{Froehly1981}. First  tests in the laboratory \citep{shaklan1987} or in an astronomical environment \citep{Connes1988} showed the potential of the technique.} The use of single-mode fibers was first motivated by their potential ability to clean the beams from the aberrations induced by atmospheric turbulence, which were the cause of poorly calibrated visibility measurements. The use of single-mode fibers for interferometry was established with the Fiber Linked Unit for Optical Recombination (FLUOR) instrument at Kitt Peak with the McMath telescope, where the ability to detect astronomical fringes was demonstrated \citep{Foresto1992}. {FLUOR was upgraded and moved to the Infrared-Optical Telescope Array (IOTA) interferometer at the Fred Lawrence Whipple observatory} where the accurate calibration of visibilities was effectively demonstrated, leading to the first astrophysical results \citep{Perrin1998}. { The use of single-mode fibers for optical and infrared interferometry has since become more common, with several instruments currently and almost in operation at the VLTI and at the Center for High Angular Resolution Astronomy (CHARA) array  \citep{LeBouquin2011, Anugu2020,  Mourard2022, Setterholm2023}}.  

However, single-mode fibers have other features that are  potentially useful for long-baseline interferometry. First, their throughput at near-infrared wavelengths can be extremely high over an octave in wavelength. The limitation in bandwidth mostly comes from the capability to keep a large fraction of the energy in the core, while the fraction in the cladding leads to losses in curved sections. This is particularly true for silica fibers, which are used for telecom applications with attenuations of as low as 0.1419\,dB/km at 1560\,nm \citep{Tamura2018}. But this is also true for fluoride-glass fibers with attenuations of as low as 1\,dB/km in the astronomical K band between 2 and 2.4\,$\mu$m \citep{Perrin2006}. Second, single-mode fibers allow us to control polarizations, while this is quite difficult or raises strong constraints with classical bulk optics with polarization rotations and birefringent effects at reflections and optical surface interfaces. A classical solution to polarization effects with fibers is to use polarization-maintaining fibers which, thanks to strongly birefringent fibers, force the orientation of linear polarization axes to remain aligned with the fiber neutral axes. {Nevertheless, a drawback of these fibers is that the {interference patterns of the two polarization axes need to be detected separately} 
unless they are synchronized with an external device; see the end of this section.} 
An alternative is to use weakly to nonbirefringent fibers and to control the polarization state using Lef\`evre loops to compensate for birefringence and polarization rotation \citep{Lefevre1980}. This principle has been successfully demonstrated  with the FLUOR beam combiner{. It was} used in the VLT INterferometer Commissioning Instrument (VINCI) \citep{Kervella2000} and improved with the Optical Hawaiian Array for Nanoradian Astronomy ('OHANA) project to reach very high fringe contrasts of close to 90\% in the full K band despite the very long fiber length (300\,m) \citep{Kotani2005}. As we show in this paper, fluoride-glass fibers have extremely weak birefringence and only the orientation of polarizations needs to be adjusted to maximize fringe contrast (Section~\ref{sec:FPR}). 

Given the ambition of GRAVITY to perform very high-accuracy astrometry on a source more than 5 magnitudes fainter than the then faintest source ever observed in optical interferometry, we made the choice to use standard { (i.e., nonpolarization-maintaining)} single-mode fluoride glass fibers feeding an integrated optics beam combiner. The fluoride-glass fibers{ indeed} have high throughput and we could contemplate a throughput as high as 90\% for the whole chain of single-mode fibers of Fig.~\ref{Fig:chain}. We had built the know-how for high throughputs with single-mode fibers with the 'OHANA project and we knew the transmission to expect for the approximately 20 meters of fiber that we thought would be necessary for GRAVITY. As there is no need to split light after standard fibers, this choice would also lead to high throughputs in particular when there is no scientific need to split polarizations to measure the Stokes parameters, thereby saving some precious percentage of throughput while the alignment of polarizations can be achieved in the fibers with fiber polarization rotators (see Section~\ref{sec:FPR}). We also knew that fluoride-glass fibers could potentially have very low birefringence, meaning two polarizations could be easily mixed without significant loss of contrast and that the weak birefringence would not prevent high-accuracy astrometry. In addition, delays can also be built inside the fibers by stretching the waveguides (Section~\ref{sec:FDDL}), allowing us to balance optical paths in double-field interferometry inside the fibers. At the end of the chain, the fibers feed the single-mode, integrated optics beamcombiner \citep{Perraut2018}. As a consequence, all the functions required to achieve the specifications of GRAVITY could {\it a priori} be made with standard fibers while ensuring higher  throughput and flexibility than with any alternative solution. This choice has been validated by the performance achieved by GRAVITY, the most sensitive interferometric instrument ever built with very high accuracy capabilities \citep{GRAVITY_1st_light_2017}.

{  Alternative concepts of fiber interferometers exist for the K band. Low-OH fibers are used for the Michigan Young STar Imager (MYSTIC) instead of fluoride glass fibers \citep{Setterholm2023} but with a much higher attenuation: 269\,dB/km compared to 4.71\,dB/km at $2.2\,\mu$m. This therefore forces MYSTIC to use as short a fiber length as possible, forbidding beam control systems such as those developed for GRAVITY. Also, birefringent plates have been proposed by \citet{Lazareff2012} to compensate for birefringence in polarization-maintaining fibers. These are used in the Precision Integrated-Optics Near-infrared Imaging ExpeRiment (PIONIER) and in the Michigan InfraRed Combiner-eXeter (MIRC-X) \citep{LeBouquin2011, Anugu2020}; they have a birefringent phase of less than $20^\circ$ and allow these instruments to 
reach contrasts of as high as 98.5\%, but with a{n extra} photometric loss of 10\%. 
\citet{Gardner2021} achieve 10$\,\mu$as astrometry {from} differential and closure phases {also using birefringence plates} with MIRC-X.}

The layout and the specifications of the fiber chain are presented in Section~\ref{sec:layout}. The characteristics of the fibers are presented in Section~\ref{sec:fibers}, in which the measurements of birefringence (a generalized Malus law with birefringence is derived in Appendix~\ref{sec:appA}), differential dispersion, and throughput are discussed. Raman scattering and fluorescence by rare earth elements caused by the metrology laser are also discussed in this section. Finally, the principles and performances of the FDDLs and fibered polarization rotators (FPRs) are presented in Sections~\ref{sec:FDDL} and \ref{sec:FPR}.
\begin{table*}[!h]
\caption[]{Main characteristics of the fibers and Fiber Control Unit
elements of GRAVITY. More details can be found in the dedicated sections of this paper.}
\label{Tab:perf}
\centering          
\begin{tabular}{ l l l }     
\hline\hline       
Item & Performance & Comment\\
\hline 
\noalign{\smallskip}
\multicolumn{3}{c}{ Fibers and FCU specifics} \\
\noalign{\smallskip}
\hline              
Operating temperature & 240\,K & tested in cryogenic environment \\
Operating pressure & $10^{-6}\,$mbar & tested in cryogenic environment \\
Core diameter & $6.55\,\mu$m & by design\\
Numerical aperture & 0.22 & at $1.9\,\mu$m\\
Cut-off wavelength & 1.88-1.9\,$\mu$m & first and second batch values (Section~\ref{sec:parameters})\\
Mode field radius ($w_0$)& 3.78-3.82\,$\mu$m & first and second batch values (Section~\ref{sec:parameters})\\
Fiber attenuation & 6.5\,dB/km@2\,$\mu$m - 3.4\,dB/km@2.5\,$\mu$m & measured with the cut-back method (Section~\ref{sec:parameters}) \\
Fiber connector losses & 0.05-0.1\,dB & with accurate alignment of cores (Section~\ref{sec:transmission}) \\
Minimum curvature radius & 2\;\!cm & by design \\
Total throughput & $\geqslant 87.5\%$ &  all fibers and connectors of Fig.~\ref{Fig:chain} (Section~\ref{sec:transmission}) \\
Birefringence beating length & 297-1110\,m & see Section~\ref{sec:birefringence}\\
Contrast loss due to birefringence & $\leqslant 6.9\%$ more likely $\leqslant 5\%$ & see Section~\ref{sec:birefringence}\\ 
Differential dispersion & $\leqslant 2.04\,\mu$rad $\,$cm$^2$ & with FDDLs at 0 stroke (Section~\ref{sec:dispersion})\\
Contrast loss due to diff. disp. & $\leqslant 0.6\%$ & with FDDLs at 0 stroke at $R=22$ (Section~\ref{sec:dispersion})\\

\hline
\noalign{\smallskip}
\multicolumn{3}{c}{ FDDL specifics} \\
\noalign{\smallskip}
\hline
Stroke of fibered delay lines & $ \geqslant 5.7\,$mm & see Section~\ref{sec:FDDL}\\
OPD difference at zero stroke & $\leqslant 2.29\,$mm & see Section~\ref{sec:FDDL} \\
Differential dispersion variation & $1\,\mu$rad $\,$cm$^2$ per mm of OPD & see Section~\ref{sec:dispersion}\\
Max contrast loss due to diff. disp. & $\leqslant 5.8\%$ & at 6\,mm stroke with $R=22$ (Section~\ref{sec:dispersion})\\
Command relative OPD accuracy & 40\,nm & see Section~\ref{sec:FDDL}\\
Relative OPD accuracy & $0.6\,$nm in 100 s & {controlled} with the GRAVITY metrology (Section~\ref{sec:FDDL})\\
\hline
\noalign{\smallskip}
\multicolumn{3}{c}{ FPR specifics} \\
\noalign{\smallskip}
\hline
Range of polarization rotators & $\geqslant 180^{\circ}$ & measured with the Malus law (Section~\ref{sec:FPR}) \\
Command accuracy & $\leqslant 1^{\circ}$ & see Section~\ref{sec:FPR} \\
\hline                  
\end{tabular}
\end{table*}
  
\section{Layout and top-level requirements}
\label{sec:layout}
The principles of the GRAVITY instrument are presented in \citet{GRAVITY_1st_light_2017}. GRAVITY primarily uses single-mode fibers to filter residual aberrations in order to achieve the high-precision interferometry described in Section~\ref{sec:introduction}. The instrument has two output channels: the Science Channel and the Fringe Tracker channel. These two channels can either be fed with two sources, a reference source for the Fringe Tracker and a science source for the Science Channel, or by a single source, that is, the science source, which then serves as both a reference and science source; in that case the light is split in the Fiber Coupler. In the former case, the GRAVITY metrology provides the difference in optical path between the two fringe pattern systems, while in the latter case the GRAVITY metrology can only be used to 
{control} moving parts in the beam such as the FDDLs (Section~\ref{sec:FDDL}). GRAVITY therefore has a total of eight interfering beams: two channels for each of the four telescopes. Each of these beams follows exactly the same design as sketched out in Fig.~\ref{Fig:chain}. Light coming from the four VLTI delay lines is focused at the entrance of the GRAVITY cryogenic vessel in the Fiber Coupler, which splits the light to feed the four injection fibers of each channel. Each fiber is connected to a FDDL (Section~\ref{sec:FDDL}) linked to an FPR (Section~\ref{sec:FPR}). The set of FDDLs and FPRs is called the Fiber Control Unit (FCU), which is used to control path lengths and polarization rotation. The FCU is then connected to the V-groove fibers feeding the integrated optics beam combiners. All connections are performed with E2000 connectors from Diamond whose specification is to ensure 0.1\,dB connections of telecom fibers.

Requirements were set on the subsystems, on the fibers (high throughput, weak birefringence, and low differential dispersion), and on the connections to achieve high sensitivity (fringe tracking at least on a K=10 source, a coherent magnitude of K=11 has been achieved \citep{Lacour2019}), high accuracy (1\% on visibility amplitudes, 0.1$^\circ$ on closure phases, and 10$\,\mu$as on astrometry), and high reliability under low pressure and cryogenic temperatures. 

The VLTI is equipped with large stroke delay lines in a 130\,m tunnel that can compensate for the  variations in optical path length that are of sidereal origin \citep{Glindemann2001}. The action of {the delay} lines 
cancels the path-length differences for a fixed point in the field but cannot compensate for path-length differences between the two targets of GRAVITY. The purpose of the FDDLs is to compensate for the residual differences, which can be as large as 6\,mm for two targets at low elevation separated by 6\,arcsec and with a differential velocity of up to {150\,nm/s} with the 200\,m maximum baseline of the VLTI using the ATs. We note that this stroke is not enough for the GRAVITY wide mode of GRAVITY+, and an additional system is required to compensate differential delays in this mode \citep{GRAVITY_Wide_2022}. Another purpose of the FDDLs is to offset the bias in optical path difference (OPD) between the fibers after minimizing differential dispersion (see Sections~\ref{sec:dispersion} and \ref{sec:FDDL}). In addition, the FDDLs need to be stable enough to get negligible OPD noise during long integrations compared to other sources, with the goal being to reach 1\,nm over 100\,s exposures. 

Because of the use of weakly birefringent standard fibers, the differential angles between the polarizations of the GRAVITY beams need to be zeroed before injection into the integrated optics chips. The requirement on the FPRs is to align the polarization axes with an accuracy of better than $8^\circ$ in order to reach contrasts of 99\% or higher. Although they are motorized, the FPRs are not used to dynamically compensate for polarization rotations in both GRAVITY and VLTI but to make GRAVITY polarization neutral. In addition, the user has the option to split and analyze polarizations with a half-wave plate downstream from the beam combiners in the spectrometers in order to measure the Stokes parameters and perform interfero-polarimetry. 

Three series of four FPRs and four FDDLs have been produced and labelled ABCD, EFGH, and IJKL. The last batch is a spare, while the first two are used in the GRAVITY instrument. ABCD are used in the Fringe Tracker channels, while EFGH are used in the Science channels. The IJKL spares have been individually characterized, but the performance (differential dispersion and throughput) of the full chain with the IJKL FDDLs and FPRs cannot be given as they have not been connected to the Fiber Coupler and V-groove fibers.

The performance of the fibers and FCU and the specific performance of the FDDLs and FPRs are given in Table~\ref{Tab:perf}. Some characteristics of the fibers and of the FCU were already presented in \citet{GRAVITY_1st_light_2017} and \citet{Perraut2018}. These are updated and discussed in more detail in the following sections.
%
%

\section{Characteristics of the fibers}
\label{sec:fibers}
%
%
%
The fibers for GRAVITY were developed by the Le Verre Fluor\'e company  in France. The class of ZBLAN glasses was discovered by \citet{Poulain1975} with promising transmission properties in the infrared. The fibers used for GRAVITY are a type of ZBLAN mostly made of zirconium (ZrF4). 

The main function of the fibers is to transport light across the instrument with minimum losses. For single-mode interferometry, another function is to filter the beams to get rid of the variations of phase due to turbulence across individual pupils. Doing so, the phase fluctuations are traded against intensity fluctuations, which are measured and calibrated out following the principles introduced for the FLUOR instrument~\citep{Foresto1997}. In the case of GRAVITY, the so-called photometric signals ---which measure the quantity of photons coupled in each fiber--- are extracted from the 24 outputs of the beam combiner by linear combination. 

{ Apart from the light pollution issue experienced with the metrology laser (Section~\ref{sec:Raman}), no major difficulty was experienced when using the fibers at 240\,K, even when stretching and twisting them to build optical delay (Section~\ref{sec:FDDL}) and to rotate polarizations (Section~\ref{sec:FPR}).} 

\subsection{Waveguide parameters}
\label{sec:parameters}
The guiding property of the fibers is obtained by slightly changing the composition of the core with respect to the cladding to get a higher index of refraction $n_{\mathrm{core}}$ compared to $n_{\mathrm{clad}}$. The fundamental mode and the number of modes carried by a step-index fiber with constant refractive indices in the core and the cladding are determined by three parameters: the numerical aperture $NA=\sqrt{n_{\mathrm{core}}^2-n_{\mathrm{clad}}^2}$, the radius of the core $a,$ and the wavelength $\lambda$; the radius of the cladding is supposed to be infinite. The fiber is single-mode for wavelengths $\lambda$ larger than the cut-off wavelength $\lambda_{\mathrm{c}}=\frac{2\pi a NA}{2.405}$~\citep{Neumann1988}. For K-band instruments in general, the aim is to set $\lambda_{\mathrm{c}}$ just below $2\,\mu$m by adjusting the other two parameters $NA$ and $a$. Light is mostly guided in the core close to the cut-off wavelength but an increasing amount of light gets guided in the cladding with increasing wavelength, causing losses if the fibers are bent. In the case of GRAVITY, the fibers also transport the metrology laser at $1.908\,$nm \citep{Rabien2008,Lippa2016} and need to be single-mode at this wavelength. Given that a large fraction of the fiber lengths are wrapped around spools in the FDDLs (15.9 to 17.5\,m out of total lengths of 20.5 to 22\,m), we chose a cut-off wavelength as close as possible to $1.908\,$nm and the fibers were therefore specified to be single-mode just below the metrology wavelength in order to minimize losses at the upper end of the K band. The fundamental mode is not a Gaussian for step-index fibers (it is a Gaussian if the index law is Gaussian across the fiber) but is a combination of Bessel functions. However, it is relatively close to a Gaussian beam and a waist is defined, the value of which is approximately: $w_0\simeq a\left(0.65+\frac{1.619}{V^{3/2}}+\frac{2.879}{V^6}\right)$, with $V=\frac{2\pi a NA}{\lambda}$ being the normalized frequency~\citep{Neumann1988}. The $f/D$ ratio of the focusing optics  is optimized to reach optimum injection in the fiber in the absence of aberrations. The maximum coupling efficiency for a uniform and full circular pupil is 78\%~\citep{Shaklan1988}. The waist varies almost linearly with the wavelength, which means that the injection efficiency is almost flat across the K band. $NA$ and $a$ are chosen so that an aberration-free focusing optics can be easily produced. Together with the constraints to be fulfilled to produce fibers of excellent quality, this led to the  choice of $NA=0.22$ and $2a=6.55\,\mu$m at 2\,$\mu$m (meaning $f/D\simeq3$ for the focusing optics) as shown in Table~\ref{Tab:perf}. With these values, the cut-off wavelength reaches 1.88 and $1.9\,\mu$m for the two batches of fibers used for GRAVITY. This was confirmed by measuring the fiber throughput for increasing curvature radius, which set the cut-off wavelength at $1.9\,\mu$m.
\begin{table*}[!t]
\caption[]{Birefringence measurements of the GRAVITY fibers at 240\,K. The test samples are cables made of a fiber from the Fiber Coupler, a FDDL, and a FPR. Roughly 2\,m of fibers at $\approx 290$\,K were connected to the polarization bench outside the cool vessel. The FDDLs and FPRs are paired (labels A to L) as used on the instrument and labeled accordingly as either Fringe Tracker or Science Channel. Telescope is the telescope number in the GRAVITY instrument. Telescope 1 is fed by either AT4 or UT4, Telescope 2 by AT3 or UT3, etc. The basic measurement is the differential birefringence phase $\delta \varphi$. No error bars are given but the method allows to detect phases as low as 0.05\,rad, giving an idea of the accuracy of the measurements. The beating length $L_{\mathrm B}$ and the minimum and maximum contrasts are deduced from this quantity. The ranges of fringe contrasts are obtained by pairing a fiber with the fibers of closest or most different measured birefringence in a given channel, the lowest contrast being obtained by adding the birefringent phases, and the highest contrasts by subtracting them.}
\label{Tab:birefringence}
\centering          
\begin{tabular}{ ccc l l l l}     
\hline\hline       
Channel & Telescope & Beam & Fiber length (m) & $\delta \varphi$ (rad) & $L_B$ (m) & Min-max contrast\,(\%)\\
\hline 
\multirow{4}{0em}{FT} & 4 & A & 18.88 & 0.35 & 339 & 93.1-100.0 \\
& 3 & B & 18.88 & 0.2 & 593 & 95.5-100.0 \\
& 2 & C & 18.88 & 0.15 & 791 & 96.2-100.0 \\
& 1 & D & 18.88 & 0.4 & 297 & 93.1-100.0 \\
\hline
\multirow{4}{0em}{SC} & 4 & E & 20 & 0.13 & 993 & 98.4-100.0 \\
& 1 & F & 20 & 0.22 & 573 & 97.4-100.0 \\ 
& 2 & G & 20 & 0.15 & 811 & 98.1-100.0 \\ 
& 3 & H & 20 & 0.24 & 531 & 97.4-100.0 \\ 
\hline
\multirow{4}{0em}{Spare} & & I & 19.43 & 0.21 & 581 & 96.9-99.9 \\ 
& & J & 19.43 & 0.29 & 421 & 96.9-99.9 \\ 
& & K & 19.43     & 0.11 & 1110 & 98.0-100.0 \\ 
& & L & 19.43 & 0.15 & 814 & 97.6-100.0 \\ 
\hline
\end{tabular}
\end{table*}
\begin{figure}[!h]
\includegraphics[width=\columnwidth]{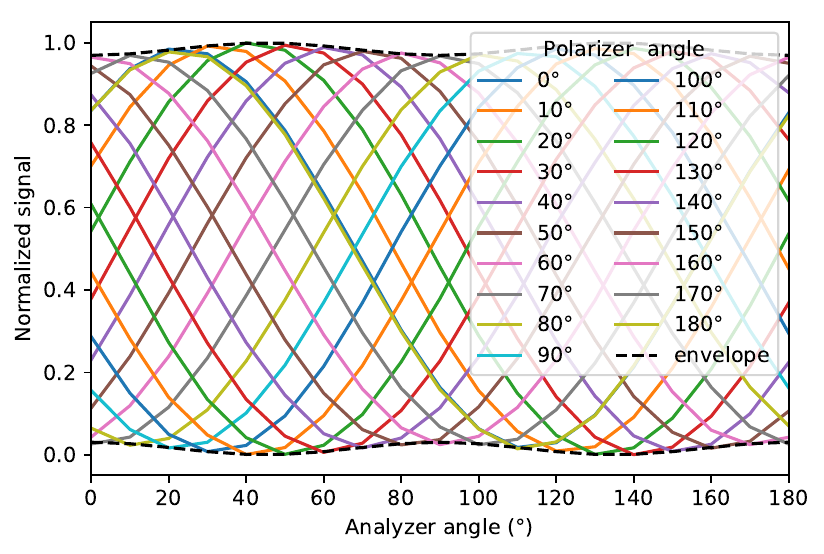}
   \caption{Example of generalized Malus laws for 18 orientations of the polarizer upstream from the injection in a chain of fibers made of FDDL~A, FPR A, and the A fiber coupler (total length of 18.88\,m), and 18 orientations of the analyzer downstream with the setup of Fig.~\ref{Fig:polarization}. The dashed line is the envelope of Eq.~\ref{fig:env} with a measured birefringence phase difference between the two polarization axes of 20$^{\circ}$, yielding a birefringence beating length of 339\,m according to the method presented in Appendix~\ref{sec:appB}.}
         \label{Fig:Malus_FPR_A}
   \end{figure}
\subsection{Birefringence}
\label{sec:birefringence}
The GRAVITY fibers were specified to be at most weakly birefringent in order to allow operation without splitting polarizations while reaching high fringe contrasts.  The aim here is to achieve  maximum sensitivity when working with faint sources and to perform astrometry between two sources with optimum accuracy.
The effect of birefringence between the two axes of polarization can be characterized by measuring the generalized Malus law of the delay lines and the {FPRs} of Section~\ref{sec:FPR} with the setup explained in Appendix~\ref{sec:appA}. The intensity for various positions of the polarizer and of the analyzer between 0 and $180^o$ is plotted in Fig.~\ref{Fig:Malus_FPR_A}.\\ \\
Assuming orthogonal neutral axes can be defined for the fiber and aligned with the $0$ or $90^{\circ}$ positions of the polarizer, following Eq.~\ref{eq:malus}, the normalized intensity (by the maximum detected intensity) measured at the output of the analyzer can be written as:
\begin{equation}
I_A(\theta_A, \theta_P,\delta\varphi) = \cos^2\left(\theta_P-\theta_A \right)-\sin^2\left(\frac{\delta \varphi}{2}\right)\sin\left(2\theta_P \right)\sin\left(2\theta_A \right)
,\end{equation} 
where $\theta_P$ is the angle of the polarizer, $\theta_A$ is the angle of the analyzer, and $\delta\varphi$ is the amount of differential birefringence phase between the two axes of polarization. 
%
%
The effect of birefringence is to reduce the amplitude depending on the angle of the polarizer in the input. There is no effect for a polarization aligned with the neutral axes of the fibers, that is, for $\theta_P=\frac{\pi}{2} (\frac{\pi}{2}),$  and the effect is maximum for a polarization rotated by $45^{\circ} (90^{\circ})$, that is, for $\theta_P=\frac{\pi}{4} (\frac{\pi}{2})$. The envelope of the family of generalized Malus laws is no longer flat, as shown in Fig.~\ref{fig:env} of Appendix~\ref{sec:appB}. This effect shows up in Fig.~\ref{Fig:Malus_FPR_A} where the envelope is plotted with the black dashed lines. This figure is compatible with the existence of neutral axes for the fiber, which in this case are aligned with the $0$ and $90^{\circ}$ positions of the polarizer. 
Equation~\ref{eq:env} of Appendix~\ref{sec:appB} gives the expressions of the upper and lower envelopes:
%
\begin{equation}
  \left\{\begin{array}{@{}l@{}}
    I_{\mathrm{low}}(\theta_A,\delta\varphi) = \frac{1}{2}-\frac{1}{2}\sqrt{1-\sin^2\left( \delta \varphi \right)\sin^2\left(2\theta_A \right)}\\
    I_{\mathrm{up}}(\theta_A,\delta\varphi) \;= \frac{1}{2}+\frac{1}{2}\sqrt{1-\sin^2\left( \delta \varphi \right)\sin^2\left(2\theta_A \right)}
  \end{array}\right.
.\end{equation}
These expressions depend on the single parameter $\delta \varphi$ and provide a method to measure the effect of birefringence. In the presence of birefringence, the maximum of the minimum is no longer zero and the minimum of the maximum is no longer 1, but these are respectively $\sin^2\left(\frac{\delta \varphi}{2}\right)$ and $1-\sin^2\left(\frac{\delta \varphi}{2}\right)$, assuming that $|\delta \varphi| \leq \frac{\pi}{2}$, that is, for a length of fiber equal to less than one-quarter of the birefringence length with a maximum modulation of the Malus laws in this particular case.
%
%
%
The difference between the propagation constants of the two polarization axes can be written as:
\begin{equation}
\Delta \beta = \frac{2\pi}{L_\mathrm{B}}
,\end{equation}
where $L_\mathrm{B}$ is the beating length of birefringence. The accumulated differential phases between the two polarizations over a length of fiber $L$ can be written as:
\begin{equation}
\delta \varphi = L\Delta\beta
,\end{equation}
and therefore for a given fiber length $L$, the beating length of birefringence can be deduced from the measurement of $\delta\varphi,$ as:
\begin{equation}
L_\mathrm{B} = L\frac{2\pi}{\delta \varphi}
.\end{equation}
The measurement of the birefringence characteristics of the fibers at 240\,K using the method presented in this paper is given in Table~\ref{Tab:birefringence}.  The beat length of birefringence was measured between 297\,m and 1110\,m for the GRAVITY fibers. 
Fringe contrasts of interferograms are  degraded by the difference in differential birefringence phase $\Delta \delta \varphi$ between the beams as they are multiplied by $\cos \left(\frac{\Delta\delta \varphi}{2}\right)$ (see e.g., \citet{Perraut1996}). As the orientations of the neutral axes of the fibers have not been controlled in GRAVITY, the worst case for the fringe contrast is obtained when the differential phases are large, have opposite signs, and add up in the above formula as $\Delta \delta \varphi = |\delta \varphi_1|+ |\delta \varphi_2|$, leading to the minimum contrasts given in the last column of Table~\ref{Tab:birefringence}. The lowest possible contrast of $93.1\%$ is achieved for the A-D baseline of the Fringe Tracker, while this value is systematically larger than $97.4\%$ for the Science Channel and larger than $96.9\%$ for the Spare. An average value between the minimum and maximum contrasts is probably more representative of the performance of GRAVITY in practice, which sets typical contrast losses due to birefringence to less than $5\%$ for the combination of the FDDLs and FPRs, which is our initial specification. Taking into account the extra lengths of ~2\,m of the Fiber Coupler and of the V-groove fibers connecting to the Integrated Optics chips to extrapolate fringe contrasts, $\Delta \delta \varphi$ needs to be increased by$~10\%,$ which reduces the minimum possible contrast to $92\%$ for the Fringe Tracker and to $97\%$ for the Science Channel with probable values of the order $95\%$ or more, 
in practice. In addition, no clear  variation of birefringence was measured during the tests when changing the FDDL extension or the FPR angle.  
\begin{table*}[!t]
\caption[]{Differential dispersion and length characteristics of the GRAVITY fiber chains at 240\,K and for zero extension of the FDDLs. No error bars are given for the differential dispersion measurements but the minimum value measured with the setup is of the order of 0.1 $\mu$rad\;\!cm$^2$, which gives an idea of the accuracy of the method. The ranges of fringe contrasts in the last column are deduced from these values and obtained by pairing a fiber with the fibers of closest or most different measured differential dispersion in a given channel. The fringe contrast is given for the lowest resolution of GRAVITY ($\lambda / \delta \lambda \simeq 22$).} 
\label{Tab:dispersion}
\centering          
\begin{tabular}{ ccc l l l }     
\hline\hline       
Channel & Telescope & Beam & Differential length (m) & Differential dispersion ($\mu$rad\;\!cm$^2$)  & Min-max contrast\,(\%)\\
\hline 
\multirow{4}{0em}{FT} & 4 & A & 0 & 0 & 99.8-100.0 \\
& 3 & B & 1.90 & 0.3 & 99.7-100.0 \\
& 2 & C & -0.27 & -1.1 & 99.7-100.0 \\
& 1 & D & 0.69 & -1.2 & 99.7-100.0 \\
\hline
\multirow{4}{0em}{SC} & 4 & E & 0 & 0 & 99.4-100.0 \\
& 1 & F & 1.48 & 2.04 & 99.4-100.0 \\
& 2 & G & -0.81 & 1.94 & 99.4-100.0 \\
& 3 & H & -0.43 & 0.35 & 99.6-100.0 \\
\hline

%
\hline
\end{tabular}
\end{table*}
%


\subsection{Differential dispersion}
\label{sec:dispersion}

Differential dispersion is taken here to mean the accumulated difference of the second-order variation of the phase of the wave as a function of wavenumber between two fibers caused by the propagation in the fibers. Namely, $\phi''(\sigma)$ if the interferogram measured by GRAVITY writes $A+B \cos \left( 2\pi\sigma x +\phi(\sigma)\right)$, where $x$ is the OPD. The differential dispersion term is measured in units of $\mu$rad\;\!cm$^2$. \\
The effect of fiber dispersion is to spread the interferogram over the OPD. More specifically, the effect of the first derivative of the phase is to shift the interferogram by $-\frac{1}{2\pi}\phi^{'}(\sigma)$ and the second-order derivative yields an OPD shift as a function of wavenumber. As a consequence, interferograms at different wavelengths will not contribute to the maximum-contrast white-light fringe at the same OPD, therefore reducing contrast in wide band. The other consequence is that measuring the fringe phase, which contains the intrinsic visibility phase information, requires a calibration. \\
Intrinsic fiber dispersion cannot be canceled but differential dispersion can be minimized. Differential dispersion has several components: material dispersion due to the dependance of refractive index on wavelength, waveguide dispersion due to the dependance of the mode on wavelength, and profile dispersion due to the variation of refractive index along the fiber \citep{Coude1995}. In addition, geometric irregularities along the fiber also add a dispersion term. The first two terms dominate and have opposite signs. Differential dispersion is classically  compensated by increasing the fiber length of one arm relative to the other in the two-arm interferometer at the expense of optical-path equalization, which is achieved with
the FDDLs in the case of GRAVITY. The differential dispersion of the 'OHANA fluoride glass-fiber cables was minimized by combining six sections of 50\,m fibers with compensating effects \citep{Kotani2005,Perrin2006}. The same technique has been applied for GRAVITY. The dispersion of each element of GRAVITY (fiber couplers, FDDLs, FPRs, and fiber bundles to the integrated optics chips) has been measured against a common reference in a Mach-Zehnder interferometer at Le Verre Fluor\'e, which was developed with LESIA in the 1990s. 
{A fused-fluoride glass-fiber coupler was used as a beam combiner for accurate and repeatable dispersion measurements.}
The various fiber components of GRAVITY  were then assembled to minimize both path length and dispersion differences, which reached a maximum of 2.17\,mm in pathlength (B-C baseline of the FT channel) and 2.04\,$\mu$rad\;\!cm$^2$  (E-F baseline of the SC channel), respectively, at zero extension for the FDDLs, which corresponds to a contrast loss of less than 1\% for the low-spectral-resolution mode of GRAVITY ($\lambda / \delta \lambda \simeq 22$), and negligible contrast losses at medium and high spectral resolutions. { This amount of dispersion would be equivalently produced by 14\,m of air in the VLTI tunnel using the model for the refractivity of air of \citet{Colavita2004}. }

 The values for each fiber chain are given in Table~\ref{Tab:dispersion}. Differential dispersion varies with the extension of FDDLs as shown in Fig.~\ref{Fig:FDDL_disp} by approximately 1\,$\mu$rad\;\!cm$^2$ per millimeter of OPD generated by the FDDLs { (equivalent to 7\,m of air dispersion per millimeter)}. In the most extreme case, where one FDDL remains at zero while another reaches the maximum stroke of 6\,mm, the maximum differential dispersion reaches 8.75\,$\mu$rad\;\!cm$^2$ (this is the case between beams E and H for example in the Science Channel) leading to an absolute fringe contrast of 94.2\%. However, this requires a source at the edge of the field at the lowest elevation and in the direction of the longest baseline with the ATs and is therefore very unlikely. In any case, this loss of 5.8\% fringe contrast would not prevent fringe tracking compared to a situation with more than 99\% fringe contrast, and the loss of contrast in the science channel at such low spectral resolution (no noticeable loss of contrast for spectral resolutions of 500 and 4000) would be easily calibrated on an unresolved target. The effect of dispersion can therefore be considered close to negligible{ on fringe contrast. Fiber dispersion needs to be calibrated for astrometry with GRAVITY. This is part of the calibration plan. The issue is minimized by using a metrology with a wavelength of 1.908\,$\mu$m very close to the K band and by measuring the dispersion of the fibers as a function of OPD every month}.

  \begin{figure}[!h]
   \includegraphics[width=8.7cm]{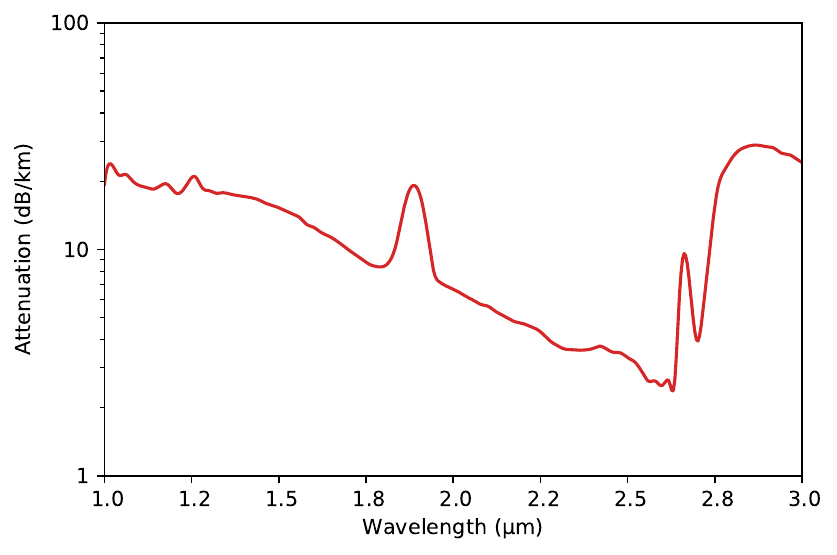}
   \caption{Attenuation of the fiber of the second batch measured by the cut-back method. The peak at $1.9\,\mu$m is due to the leakage of high-order modes as the fiber becomes single mode.}
              \label{Fig:attenuation}%
    \end{figure}
\begin{table*}[!t]
\caption[]{Throughput of the GRAVITY fiber chains. Each chain comprises the Fiber Coupler fiber, an FDDL, an FPR, and a V-groove fiber. The accuracy on the throughput measurements is of the order of 0.1\%}
\label{Tab:transmission}
\centering          
\begin{tabular}{l c c c c |c c c c }     
\hline\hline       
Channel & \multicolumn{4}{c|}{FT} & \multicolumn{4}{|c}{SC} \\ 
\hline 
Beam & A & B & C & D & E & F & G & H \\
Loss (dB) & 0.58 & 0.32 & 0.47 & 0.38 & 0.45 & 0.43 & 0.50 & 0.46 \\ 
Throughput (\%) & 87.5 & 92.9 & 89.7 &  91.6 & 90.2 & 90.6 & 89.1 & 89.9 \\
\hline
\end{tabular}
\end{table*}

\subsection{Throughput}
\label{sec:transmission}
The throughput of the fiber chains is determined by several factors; first of all, by the intrinsic transparency of fluoride glass fibers. The attenuation by the fibers used for GRAVITY is shown in Fig.~\ref{Fig:attenuation} as measured with the cut-back method on the last batch of fibers produced for GRAVITY. The attenuation increases towards the blue because of Rayleigh scattering and increases after 2.3$\,\mu$m because of absorption by water. The attenuation has a local peak at 1.9$\,\mu$m because of the suppression of high-order modes as the fiber becomes single mode. The attenuation is 6.63\,dB/km at 2$\,\mu$m and 3.65\,dB/km at 2.4$\,\mu$m, yielding transmissions of respectively 97\% and 98\% for 20 meters of fiber. \\
Connections are another possible source of transmission loss. Photons are lost at each of the three connections (Fig.~\ref{Fig:chain}) between the Fiber Coupler fiber, the FDDL, the FPR, and the V-groove fiber. Part of the loss is due to Fresnel reflection at the air--glass interface, while another part is due to the misalignment of fiber cores. We used E2000 PC connectors from Diamond to minimize both effects. The two fiber ends are held in contact with each other in the adapter to minimize Fresnel losses. The fibers are held in a ceramic ferrule encapsulated in a soft metal jacket. The soft metal jacket allows the core-to-core alignment between paired fiber ends to be adjusted very accurately. An accuracy of $1\,\mu$m is reached with maximum losses of 0.05 to 0.1 dB per connection or a total additional loss of 3.4\% to 6.7\% per fiber chain. \\
The third potential main factor is the bending of the fibers in the FDDLs. Losses are all the more important as the wavelength is large as the wave is less confined in the core of the fiber. A first theory was put forward by \citet{Marcuse1976a} assuming infinite radius for the cladding. A more realistic approach has been proposed to account for the limited radius of the cladding \citep{Faustini1997}, even with metallic claddings \citep{Peng2017}. The bend loss in dBs for a given curvature radius $R$ and length of bend $L$ can be written (adapted from e.g., \citet{Peng2017}): 
\begin{equation}
L_s = 10 \log_{10} \left[ \exp \left( 2 \alpha a \frac{L}{a} \right) \right]
,\end{equation}
where $a$ is the radius of the fiber core and $2\alpha$ is the bend-loss factor. \citet{Marcuse1976b} computed $2\alpha a$ as a function of the normalized radius of curvature $R/a$ for a single-mode fiber at the cutoff frequency for values ranging between 300 and 5000. The exponential law can be extrapolated to $R/a=6154$ in the case of the GRAVITY fibers yielding a bend loss of $6\times10^{-28}$ dB for 10 meters of bend length, which is {a} completely negligible loss. Although the mode is 25\% larger at the upper edge of the K band at 2.4\,$\mu$m, the same conclusion applies at this wavelength.\\
In conclusion, the losses are in practice dominated by the connections and this is most certainly what the results of Table~\ref{Tab:transmission} show with a minimum throughput of 87.5\% and an average throughput of 90\%, which is excellent given the number of functions performed by the fibers: namely beam transport, polarization control, differential delays, and feeding of the beam combiners.

\subsection{Raman scattering and fluorescence}
\label{sec:Raman}
GRAVITY uses a 1\,W laser at 1908\,nm to measure the OPDs between the science and reference beams in each of the four telescopes. The principles of the metrology system are described in \citet{Lippa2016}. A small fraction (less than 1\%) of the laser light is injected backwards from the two beam combination points and is propagated upward to the telescope entrance pupils to have a full measurement of the OPDs between the science and reference beams in all four telescopes. The high-power fraction (more than 99\%) of the laser beam is overlaid with the faint fraction at the entrance of the fiber couplers where the beams coming from the telescopes are fed into the fibers. This three-beam scheme was chosen to minimize the amount of power injected into the single-mode waveguides of GRAVITY. In the initial setup, only the laser beams going through the beam combiners were used, hence with full power going through the fibers. This led to high levels of light back-scattered into the GRAVITY spectrometers because of two distinctive effects that produce additional photon noise in the K band. These effects are described in \citet{Lippa2018}. One is Raman scattering, which is the consequence of the interaction between a photon and a phonon. In one configuration of the interaction, a laser photon loses energy to a phonon and is shifted to a wavelength of around 2150\,nm in the K band. The other effect is fluorescence of the rare earth elements holmium (Ho$^{3+}$) and thulium (Tm$^{3+}$), which are used to produce doped fibers. The GRAVITY fibers were contaminated at the ppm level by these elements, which is enough to produce faint signals in the K band up to 2100\,nm by excitation of the holmium ions with the very bright metrology laser. The combination of the two effects leads to contamination of the blue part of the K band up to 2200\,nm (see Fig. 4 of  \citep{Lippa2018}). A second batch of fibers was produced with reduced contamination, which allowed minimization of the contamination by fluorescence but had no effect on the Raman scattering part. The three-beam scheme allows the effects to be mitigated, with the remaining backscattering still being slightly larger than the thermal background, but only by a factor of a few and only for a few pixels at the shortest wavelengths in the detector band.


\section{Fibered differential delay lines}
\label{sec:FDDL}
\begin{figure}[t]
\includegraphics[width=\columnwidth]{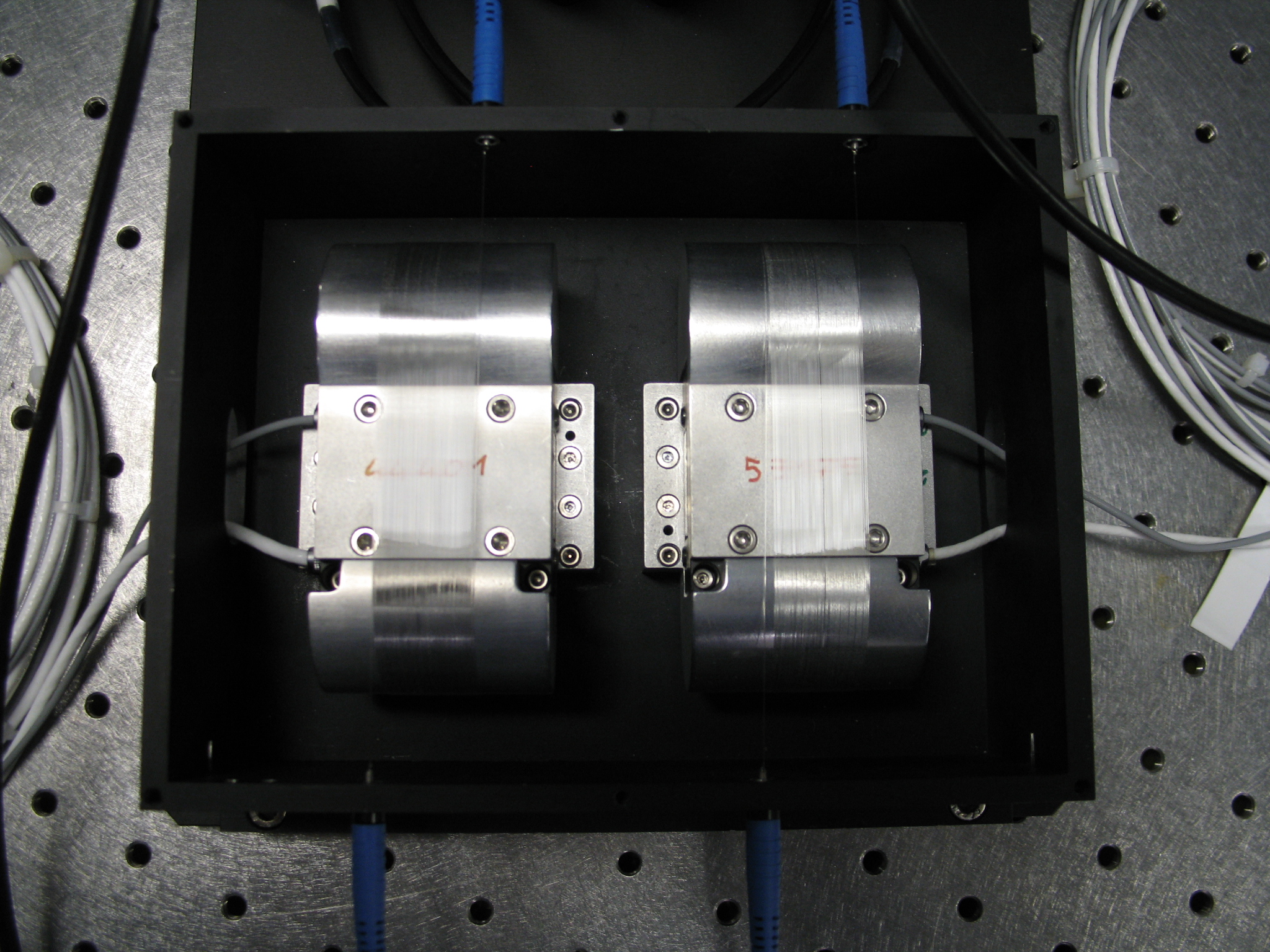}
   \caption{Pair of GRAVITY FDDLs. {Four} blue fiber jackets are visible at the top and bottom of the box. The fibers are wrapped on two half cylinders, one fixed and one mounted on a piezo actuator. Delay is produced by stretching the fibers with the piezos. The gray and white cables on the left and right are the power and gauge sensor cables of the piezos.}
         \label{Fig:FDDL}
   \end{figure}
GRAVITY has a moderate need for differential delay compensation. The maximum differential delay is for the use of the ATs for which a maximum baseline of 200\,m can be reached. The distance between the science and reference targets of 6 arcsec is also maximum with the ATs. The two combined lead to a maximum delay of 6\,mm and a maximum differential speed of the zero OPD of 150\,nm/s at the {lowest} elevation. Such static and dynamic delays can be obtained by stretching fibers. All GRAVITY beams are equipped with FDDLs and the maximum differential delay between the reference and the science targets is consequently twice the stroke of {the} FDDLs. A stroke of 3\,mm per FDDL would therefore be sufficient. However, the minimization of differential dispersion between slightly inhomogeneous fibers leads to differential lengths, which need to be compensated for. We therefore set a goal to reach 6\,mm maximum stroke for each FDDL in order to have some margin.\\
The first attempts at {fibered delay lines} were made by the IRCOM/XLIM group at Universit\'e de Limoges in the 1990s with silica fibers. The technique was  used to produce a linear optical path modulation, first with a moderate stroke of $20\,\lambda$  at an accuracy of $\lambda / 200$ ($\lambda = 633\,$nm) with a laser {control} system~\citep{Reynaud1993b}.  \citet{Zhao1994} then developed a fibered Mach-Zehnder interferometer at Paris Observatory with standard fluoride-glass fibers, whose optical path was scanned with a fiber wrapped on a piezo cylinder. The stretching of curved fibers introduces birefringence and differential dispersion, both of which have been analyzed \citep{Zhao1995a,Zhao1995b}. All these lead to moderate delays of a few thousand $\lambda$ at most. \citet{Simohamed1996} demonstrated a 318\,mm stroke with 20\,m of stretched polarization maintaining fibers. {Fibers were positioned} in 90$^{\circ}$ V-grooves with  an isotropic distribution of stress in the fiber core in order to minimize induced birefringence. {\citet{Simohamed1997}} reached a 2\,m stroke with 100\,m of fibers wrapped on an expanding cylinder. \\ 
The relative stretch of a 125$\,\mu$m fluoride glass was measured to be 0.1\% or 2$\,\mu$m/(mm.N), which therefore allows us to generate several millimeters of delay with sufficient fiber length and sufficient force applied to the fiber. 
Given the stringent accuracy requirements of GRAVITY, the challenges in building the delay lines were related to the throughput, in particular the potential losses due to the curvature of fibers in the spools, the control accuracy, the birefringence, and the second-order phase dispersion variations as a function of stretch. The characteristics of the FDDLs are listed in Table~\ref{Tab:perf}.
\begin{figure}[!h]
\begin{tabular}{c}
\includegraphics[width=\columnwidth]{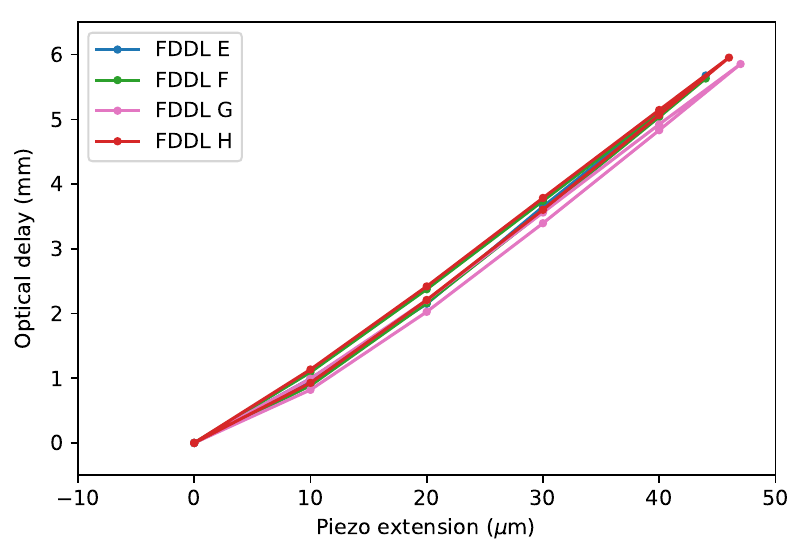} \\\includegraphics[width=\columnwidth]{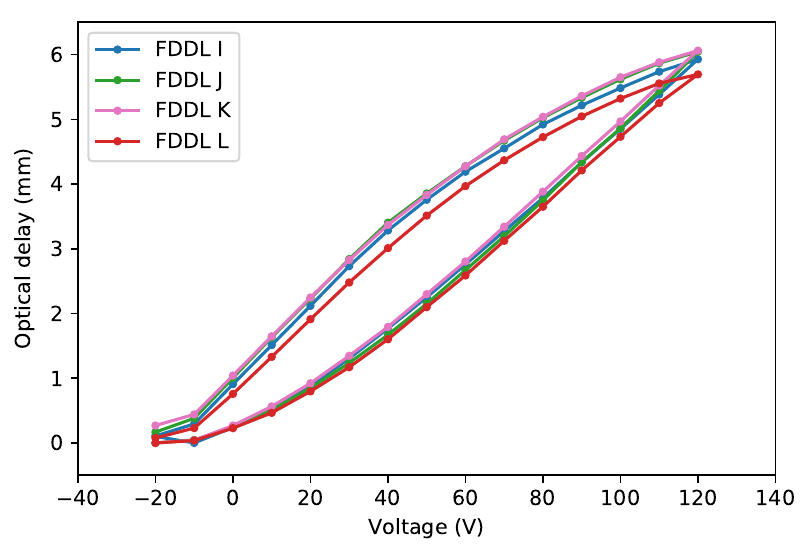} 
\end{tabular}
   \caption{Examples of optical path-length laws for two batches of Fibered Differential Delay Lines (FDDL). FDDL E, F, G, H were built with the first generation of GRAVITY fibers while I, J, K, L were built with the second generation of GRAVITY fibers. An optical delay of up to 6\,mm is built in both cases. The command is the piezo extension for E, F, G, H while it is a voltage for I, J, K, L to show the amount of hysteresis corrected by the Physik Instrument E-621 servo controller.}
         \label{Fig:FDDL_stroke}
   \end{figure}
\begin{figure}[!h]
\begin{tabular}{c}
\includegraphics[width=\columnwidth]{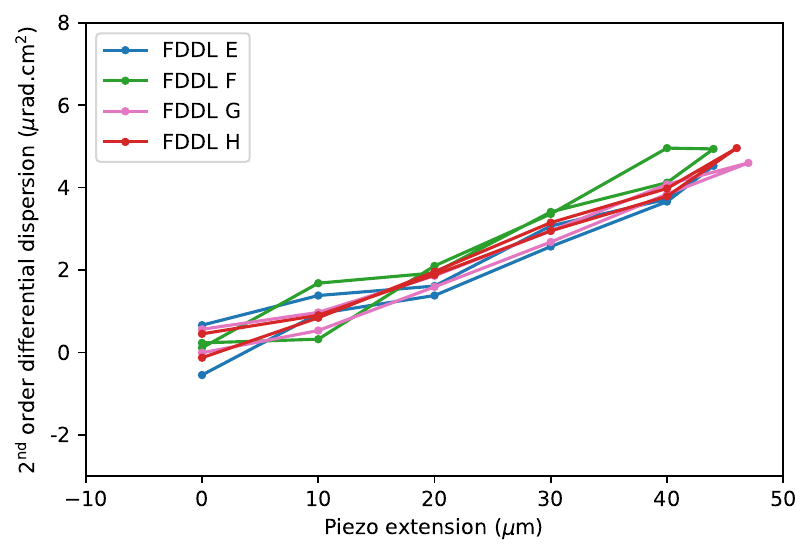} \\\includegraphics[width=\columnwidth]{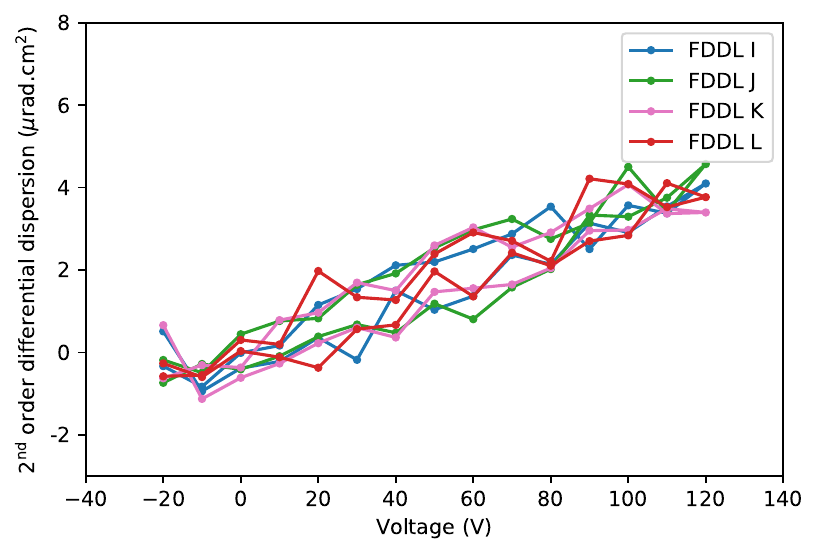} 
\end{tabular}
   \caption{Examples of second-order differential dispersion laws for the two sets EFGH and IJKL of FDDLs. The average dispersion was set to zero at minimum FDDL extension. Both sets of delay lines generate a maximum of $5\,\mu\mathrm{rad}.\mathrm{cm}^2$ at a maximum extension of 6\,mm. Abscissas are as in Fig.~\ref{Fig:FDDL_stroke}.}
         \label{Fig:FDDL_disp}
   \end{figure}
\\
The issue of throughput and curvature radius for the spools is addressed
in Section~\ref{sec:transmission}. With a 4\;\!cm spool diameter, no significant
loss is to be expected. \\
%
{The high sensitivity of fibers to mechanical and thermal stresses is a characteristic exploited to build sensors with fiber optics}
{and this sensitivity to physical conditions is an issue when using them for outdoor interferometry, as in aperture synthesis arrays.}
Fibered delay lines were for example used in 'OHANA primarily as  dispersion compensators  as the {lengths of the} fibers were not {controlled with a metrology system} 
and suffered from temperature fluctuations along the 300\,m of fiber length \citep{Vergnole2004,Kotani2005}. Following a first experiment by \citet{Connes1988}, \citet{Reynaud1992} then showed how to stabilize silica fiber lengths with a laser metrology in order to build a fiber interferometer. A similar technique {is} used for GRAVITY. To fulfill the specifications on path-length accuracy and time response, the delay lines were designed as two-stage devices: the fibers are wrapped on half-cylinders whose relative distances are actuated by translation stages with servo {controlled} piezo stacks (modified P-611.1S translation stages from Physik Instrument with 100$\,\mu$m stroke; the control of the piezos is operated with an E-621 controller, the piezo length being measured with a strain gauge sensor) and the overall path-length stability is controlled by the GRAVITY metrology coupled with the GRAVITY piston actuators. This second stage is necessary because of the intrinsic hysteresis of the spools of fibers wrapped around the half cylinders. The motion of the piezo is mechanically amplified by a pantograph whose elasticity is constrained by the loops of fibers. The effect of the expansion of the piezo applied to the pantograph is therefore not linear because of the resistance of the fibers, and the final stroke of the FDDLs is not as large as what is theoretically attainable with the piezo stacks. 
The control electronics of the FDDLs were designed to ensure that the error of FDDL commands have negligible impact on both fringe contrast and OPD error measurement for typical integration times of several minutes. 
{This requires at least 18-bit precision. For 6\,mm of range this translates to a precision of 40\,nm per FDDL. For long exposure fringes this translates to  visibilities of 99\% and for OPD measurements to accuracies of 0.6\,nm.}
We designed a real-time controller providing 20-bit precision commands (6\,nm precision on a 6\,mm range) so that the final performance of our architecture is limited by the performance of the Physik Instrument amplifier. The amplifier has an internal noise of 800~$\mu$V and requires 18-bit coding to keep the command noise to 460~$\mu$V for a voltage of between 0 and 120 V (some measurements by PI showed down to 100~$\mu$V or 20-bit precision noise in closed loop). In addition, the amplifier has a 100 kHz bandpass, which is far beyond the requirements for GRAVITY given the very slow differential speed between two channels. \\
The half cylinders were manufactured with perfectly smooth surfaces and without any grooves. The fibers were wrapped on the half cylinders with a constant tension provided by a 25\,g weight attached to a fiber end to generate a tension of 0.25\,N. The wrapping was performed without any torsion of the fiber in order to minimize constraints. As a result, the measured birefringence is very low (see Section~\ref{sec:birefringence}) and no noticeable variation of birefringence with FDDL extension has been measured.
Examples of optical path length generation by FDDLs as a function of piezo extension or of  voltage applied to the piezo are shown in Fig.~\ref{Fig:FDDL_stroke}, with a maximum delay of 6\,mm. The E-L delay lines are shown here, while a total of 12 have been produced and labelled A to L. The A-D and E-H were produced first and the I-L were produced later as spares, the difference with the 8 first delay lines being essentially the lowest contamination of the fibers by holmium and thulium. The minimum stroke of 5.7\,mm obtained for the F or L delay lines leaves a sufficient margin with which to compensate the maximum bias in OPD of 2.29\,mm between beams F and G (Table~\ref{Tab:dispersion}). \\
Figure~\ref{Fig:FDDL_disp} shows the amount of second-order dispersion generated by stretching the delay lines of Fig.~\ref{Fig:FDDL_stroke}. The second-order dispersion varies for the two batches of FDDLs by approximately  $1\,\mu\mathrm{rad}\,\mathrm{cm}^2$ per mm of generated delay. This additional dispersion reduces the fringe contrast by 5.8\% in the most extreme case with a dispersion of $R\simeq 22$ in the K band, as seen in Section~\ref{sec:dispersion}. In practice, this has a negligible effect on the capability of GRAVITY to fringe track and can be easily calibrated.  \\
%

%
%

%
\section{Fibered polarization rotators}
\label{sec:FPR}
\begin{figure}[!h]
\includegraphics[width=\columnwidth]{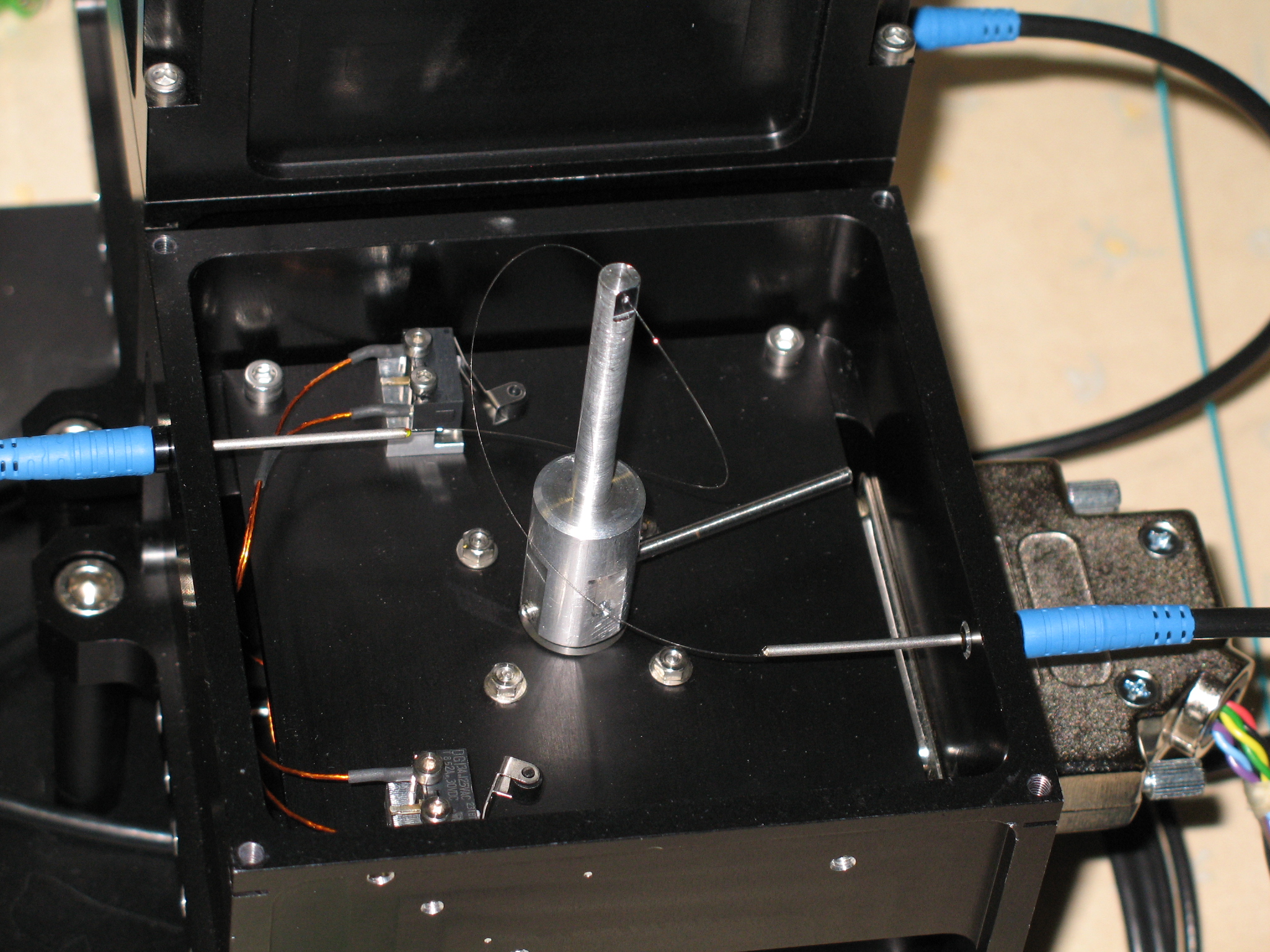}
   \caption{GRAVITY fibered polarization rotator. The blue jackets are the fiber output and input. The loop of fiber goes through a hole at the top of the shaft, which defines the rotation axis. The rotation of the shaft is controlled by a stepper motor. The two mechanical end switches are visible at the bottom of the box. The torsion applied to the fiber by the rotation of the shaft generates a rotation of the polarization of the beam going through the fiber.}
         \label{Fig:FPR}
   \end{figure}
We made the choice to not use polarization-maintaining fibers for GRAVITY, but to use weak birefringence fibers instead, allowing us to avoid systematic splitting of polarizations and therefore to optimize sensitivity and  ensure that GRAVITY is as polarization neutral as possible. The prerequisite is the very low birefringence of the fluoride-glass fibers, which was demonstrated with the very demanding 'OHANA project following the achievements with FLUOR and measured in the case of the GRAVITY fibers (Section~\ref{sec:birefringence}). As the GRAVITY fibers do not maintain polarizations, {the axes of linearly polarized} radiation can rotate after propagation in the fibers. For weakly birefringent fibers, polarization neutrality can only be achieved if the differential orientation of the axes of polarization can be canceled. The traditional way to align polarization axes in fibers is to use Lef\`evre loops \citep{Lefevre1980}. The basic principle of the device is to induce birefringence in the fibers through stress by bending the fibers. The device is equivalent to wave plates that allow polarizations in the fibers to be controlled. Such devices have been used for FLUOR in a simplified version with two loops only in two perpendicular planes to rotate polarizations without introducing birefringence, as it was found that birefringence was not the primary cause of contrast loss in interferometers with standard fluoride-glass fibers \citep{Perrin1998}. The system was further simplified by Le Verre Fluor\'e, who later showed that a single loop was{ sufficient}. Such a simplified version was used for FLUOR, VINCI \citep{Kervella2000}, and 'OHANA \citep{Perrin2006}. This is the basis of the system used for GRAVITY, which is described in this section. \\
\begin{figure}[!h]
\begin{tabular}{c}
\includegraphics[width=\columnwidth]{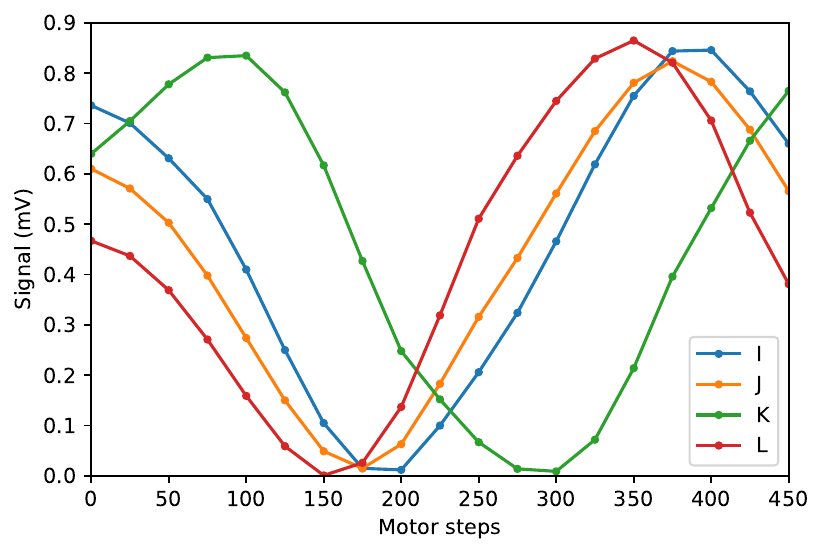} \\\includegraphics[width=\columnwidth]{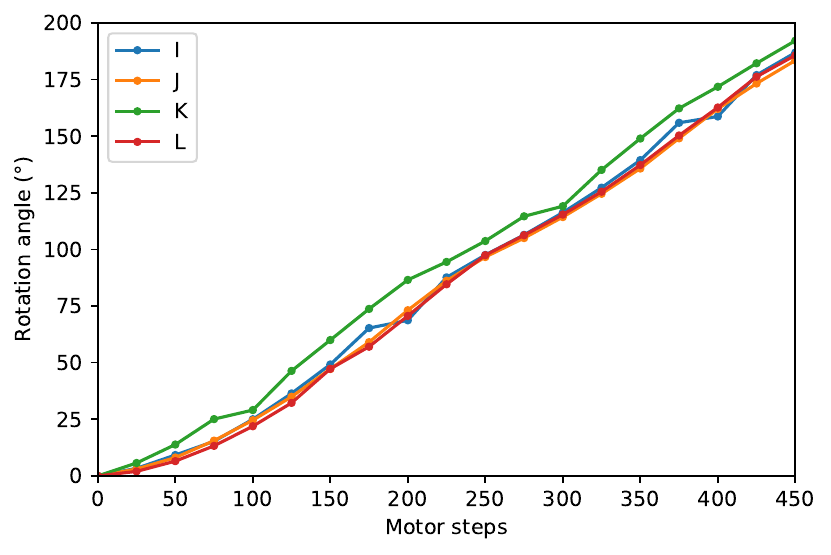} 
\end{tabular}
   \caption{Examples of Malus laws (top) and angle laws (bottom) for the fibered polarization rotators IJKL. The angle has been arbitrarily set to $0^{o}$ for 0 motor steps.} 
         \label{Fig:FPR_angle}
   \end{figure}
The principle of the GRAVITY FPR is quite simple: a torsion is applied to a loop of fiber across a diameter in order to rotate polarizations. The curvature radius is sufficiently large to avoid losses and birefringence induced by stress. An example of FPR is presented in Fig.~\ref{Fig:FPR}. A shaft is actuated by a stepper motor to twist the fiber along the vertical axis. The fiber is loose in the hole at the top of the shaft and at the input and output feedthroughs. End switches allow the twist range to be controlled in order to avoid damaging the fiber. \\
The actuator is a Phytron stepper motor VSS-UHVC 25.200.0.3 compatible with the temperature of operation. The torque required to twist the fiber loop is 0.035 mN.m, while the stepper motor yields a 12 mN.m torque. The stepper motors are driven by a Programmable Logic Controller from Beckhoff. Motion commands are sent in units of micro-steps: eight micro-steps per real motor step. One full motor revolution is accomplished in 200 steps so that a full revolution requires 1600 micro-steps. This provides ample resolution to accurately rotate the polarization by $0.25^{\circ}$ per micro-step. The motor is turned off when not in use by setting the hold current to zero to avoid heating in vacuum and errors due to fluctuating power. The FPRs lie in a horizontal plane so that no mechanical motion can be set by gravitation. \\
The performance of the FPRs was measured at 240\,K. Malus laws were measured by injecting linearly polarized light into the FPRs and detecting the rotated polarization with an analyzer at the output. Examples of resulting Malus laws are presented in Fig.~\ref{Fig:FPR_angle}. The rotation angle versus motor step command laws can be easily deduced from the Malus laws and are presented in the same figure. From these, the goal is achieved with an amplitude of rotation of larger than $180^{\circ}$. The amplitude could still be increased by moving the end switches slightly. The rotations are reproducible with an accuracy of the order of $1^{\circ}$ or better ($8^{\circ}$ is necessary to reach  99\% fringe contrast). In practice, the FPRs are used to align the linearly polarized light of an internal calibration source \citep{GRAVITY_polarization_2023} which ensures a high level of visibility contrast on sky \citep{GRAVITY_1st_light_2017}. The FPRs are not operated during the observation nights and remain at fixed positions.




%
%
%

\section{Conclusions}
The interest in single-mode fibers for high-precision interferometry has become firmly established since the concept was introduced in the early 1980s. Later developments showed that fibers are also efficient for beam transportation. In this paper, we present the characteristics of the single-mode fibers and associated functions designed for the very demanding GRAVITY instrument. Very weakly birefringent standard fibers are used to maximize  both the sensitivity of the instrument and the accuracy of astrometric measurements. Fibers are stretched in the FDDLs to compensate for differential delays of up to 6\,mm between the fringe patterns in the two fields of GRAVITY. Fibers are twisted in the FPRs to align polarizations between the GRAVITY arms, with the possibility to compensate for up to $180^{\circ}$ of differential rotation in order to maximize fringe contrast. The typical contrast losses due to differential birefringence are of the order of $5\%$ while the contrast losses due to differential second-order dispersion are less than $1\%$ in the lowest spectral resolution mode of GRAVITY. These latter would reach a maximum of less than $6\%$ in the extreme case of a science source at the edge of the theoretical field of view of 6" on the longest baseline of the VLTI with the ATs and at the lowest elevation possible. The contrast losses would be completely negligible with the medium- and high-resolution modes in the science channel. In addition to the fantastic coherence and polarimetric performance of the GRAVITY fibers, the throughput for the 20.5-22\,m chains is very high and ranges between 87.5\% and 92.9\%, with connection losses included. Overall, the performance of the GRAVITY fibers and their functions meet the desired specifications or exceed them. { There is no equivalent of this fiber system with which to route and modally filter beams -- with delay and polarization control -- in any other K-band beamcombiner.}


\appendix

\renewcommand{\thefigure}{A\arabic{figure}} 
\begin{figure*}[h]
\centering
\includegraphics[width=15cm]{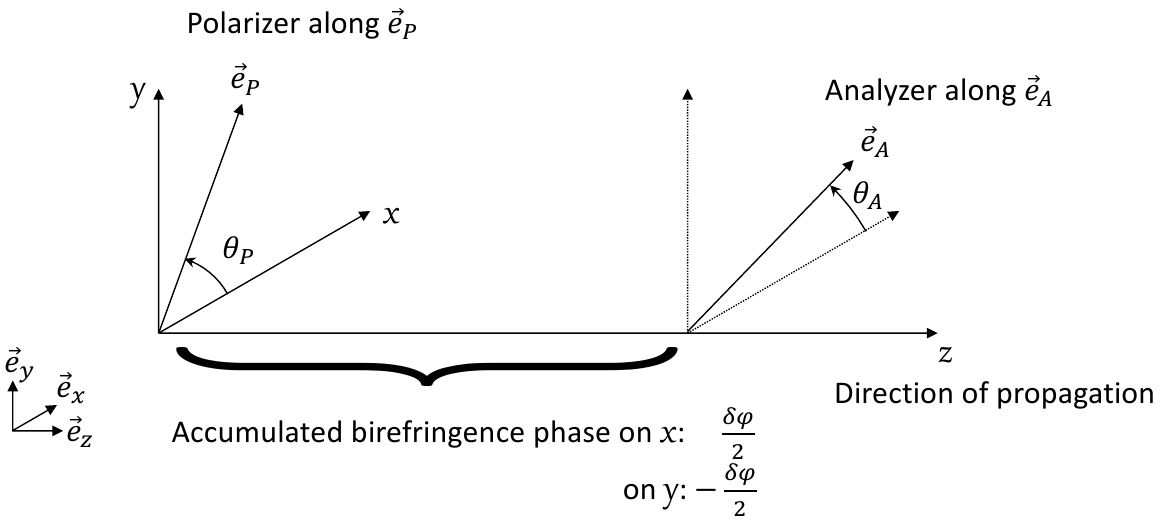}
\caption{Setup used to measure the generalized Malus laws of a fibered device as a function of input linear polarizer angle $\theta_P$ and output analyzer angle $\theta_A$.  }
\label{Fig:polarization}
\end{figure*}
\section{Generalized Malus law}
\label{sec:appA}

The geometry and notations of the setup to measure the Malus law along the axis of the fiber are given in Fig.~\ref{Fig:polarization}. 
The wave used to analyze the properties of the fibers is described by a complex vector $\overrightarrow{E}(z,t)$. It propagates along the axis $\overrightarrow{e_z}$. Here, $\overrightarrow{E}$ is written at a given distance $z$:
\begin{equation}
\overrightarrow{E}(z,t)=e^{-i(kz-\omega t)}\left( E_x\overrightarrow{e_x} + E_y \,\overrightarrow{e_y} \right)
.\end{equation}
For an unpolarized wave, $\left< E_x^2\right>=\left< E_y^2\right>=\frac{1}{2}I_{np}$. For a polarized wave,
\begin{equation}
\begin{bmatrix}
E_x \\
E_y
\end{bmatrix}
= E_p
\begin{bmatrix}
\cos \psi e^{-i\varphi_x} \\
\sin \psi e^{-i\varphi_y}
\end{bmatrix}
,\end{equation}
with $E_p^2=I_p$.

Here, $\theta_P$ is the orientation of the polarizer axis with respect to $\overrightarrow{e_x}$ counted positively towards $\overrightarrow{e_y}$. 
Let us now assume that the fiber is birefringent with $\overrightarrow{e_x}$ and $\overrightarrow{e_y}$ , which are the neutral axes of the fiber, with $\frac{\delta \varphi}{2}$ being the amount of birefringence phase accumulated along $\overrightarrow{e_z}$ on the $\overrightarrow{e_x}$ axis and $-\frac{\delta \varphi}{2}$ being the amount of birefringence phase accumulated on the $\overrightarrow{e_y}$ axis (time has been shifted to have opposite amounts of birefringent phase on the $x$ and $y$ axes). \\ 
After the polarizer, the wave is projected on the axis of the polarizer $\overrightarrow{e_P}$ and is {written as:}
\begin{equation}
\overrightarrow{E_P}(\theta_P,z,t)=e^{-i(kz-\omega t)} \left( \cos \theta_P E_x+  \sin \theta_P E_y \right) \overrightarrow{e_P} 
.\end{equation}
The wave is then injected into the fiber and propagates down to the analyzer (we assume the coupling to the fiber is lossless). The wave accumulates birefringent phases on the $x$ and $y$ axes. At the fiber output, this can be written as:
\begin{eqnarray}
\overrightarrow{E'_P}(\theta_P,z,\delta\varphi,t) &=& e^{-i(kz-\omega t)} \left( \cos \theta_P E_x+  \sin \theta_P E_y \right) \times \\ \nonumber
& &
\left( e^{-i\frac{\delta \varphi}{2}}  \cos \theta_P \,\overrightarrow{e_x} + e^{i\frac{\delta \varphi}{2}}\sin \theta_P \,\overrightarrow{e_y} \right). 
\end{eqnarray}
 The wave is filtered at the output of the fiber with an analyzer rotated by an angle $\theta_A$ with respect to the $\overrightarrow{e_x}$ axis. The effect of the analyzer is to project the incoming wave in the direction $\overrightarrow{e_A}=\cos \theta_A \,\overrightarrow{e_x} + \sin \theta_A \,\overrightarrow{e_y}$, giving the analyzed wave:
\begin{eqnarray}
\overrightarrow{E_A}(\theta_A,\theta_P,z,\delta\varphi,t) &= & e^{-i(kz-\omega t)} \left( \cos \theta_P E_x+  \sin \theta_P E_y \right) \times \\ \nonumber
& & \!\!\!\!\!  \!\!\!\!\!  \left( e^{-i\frac{\delta \varphi}{2}}  \cos \theta_A\cos \theta_P  + e^{i\frac{\delta \varphi}{2}}\sin \theta_A \sin \theta_P \right) \,\overrightarrow{e_A}. 
\end{eqnarray} 
%
The average intensity detected at the output of the analyzer is independent of $z$ and $t$ and is written as:
\begin{eqnarray}
I_A(\theta_A,\theta_P,\delta\varphi) &= &  \left< \left| \cos \theta_P E_x+  \sin \theta_P E_y \right|^2\right> \times \\ \nonumber
& &  \left| e^{-i\frac{\delta \varphi}{2}}  \cos \theta_A\cos \theta_P  + e^{i\frac{\delta \varphi}{2}}\sin \theta_A \sin \theta_P \right|^2.
\end{eqnarray} 
The first term takes different values depending on the nature of the input wave; for an unpolarized wave, it is equal to $\frac{1}{2}I_{np}$, while for a polarized wave it is equal to $I_p \cos^2\left( \psi - \theta_P \right)$. This is the classical Malus law. Developing and reorganizing the second term, for the intensity one obtains:
\begin{eqnarray}
\label{eq:malus} 
 I_A(\theta_A,\theta_P,\delta\varphi) &=& \\ \nonumber
 & & \!\!\!\!\! \!\!\!\!\! \!\!\!\!\! \!\!\!\!\! \!\!\!\!\! \!\!\!\!\! \!\!\!\!\! \!\!\!\!\! \!\!\!\!\! \begin{cases}
    \frac{1}{2}I_{np} \left[ \cos^2\left( \theta_A - \theta_P\right)   \right.- \left. \sin^2\left(\frac{\delta \varphi}{2}\right) \sin(2\theta_A)\sin(2\theta_P) \right] \\
    I_p \cos^2\left( \psi - \theta_P \right) \left[ \cos^2\left( \theta_A - \theta_P\right)   \right.- \left. \sin^2\left(\frac{\delta \varphi}{2}\right) \sin(2\theta_A)\sin(2\theta_P) \right], 
  \end{cases}
\end{eqnarray}
where the first line is the unpolarized case and the second one is the polarized case. The {cosine} term in {the} square parentheses 
is the classical Malus law for polarized light and the second term is the effect of birefringence. We note that in the case of polarized light, there is a cascade of two Malus laws as the light filtered by the two polarizers is already polarized in the input.  

\section{Measurement of birefringence}
\label{sec:appB}
\renewcommand{\thefigure}{B\arabic{figure}} 

We show in Appendix~\ref{sec:appA} that the classical expression of the Malus law needs to be modified to account for birefringence. Conversely, it can be used to measure the birefringence properties of the fiber. Natural light is  used to measure the properties of the GRAVITY fibers. 
In the following, we therefore consider the first line of Eq.~\ref{eq:malus} for unpolarized light. It is a family of curves with parameters either $\theta_A$ or $\theta_P$ with  $\theta_P$ or $\theta_A$ being the respective variable. We show here that the characteristics of the envelope of this family depend on the birefringence phase $\delta \varphi$.  \\ 

In the following, we consider $\theta_P$ as the parameter and $\theta_A$ as the variable. We highlight the fact that Eq.~\ref{eq:malus} is symmetric and therefore either case can be chosen. The expression of the envelope of the curves is classically deduced by expressing that it both takes generalized Malus law values and is tangent to the curves. It amounts to solving the system of equations:
\begin{equation}
  \left\{\begin{array}{@{}l@{}}
    I_A(\theta_A, \theta_P,\delta\varphi) - I_{\mathrm{env}}(\theta_A,\delta\varphi) =0\\
    \frac{\partial}{\partial \theta_P} (I_A(\theta_P, \theta_A,\delta\varphi) - I_{\mathrm{env}}(\theta_A,\delta\varphi) ) = 0
  \end{array}\right.
\end{equation}
by eliminating $\theta_P$ and getting the expression of $I_{\mathrm{env}}$. The system of equations is easier to solve by rewriting Eq.~\ref{eq:malus} as the real part of a complex number:
\begin{eqnarray}
I_A(\theta_A, \theta_P,\delta\varphi) &=& \frac{I_{np}}{4}\times\mathrm{Re}\left[ \cos^2\left(\frac{\delta \varphi}{2}\right) \left[ 1+ e^{2i(\theta_A - \theta_P)} \right] \right. \\ \nonumber
& & \;\;\;\; \;\;\;\; \;\;+ \left. \sin^2\left(\frac{\delta \varphi}{2}\right) \left[ 1+ e^{2i(\theta_A + \theta_P)} \right]\right].
\end{eqnarray}
Defining $c_A=\cos(2\theta_A)$, $s_A=\sin(2\theta_A)$, $c_P=\cos(2\theta_P)$, $s_P=\sin(2\theta_P)$, $\gamma^2=\cos^2\left(\frac{\delta\varphi}{2}\right)$, and $\sigma^2=\sin^2\left(\frac{\delta\varphi}{2}\right)$, the above system can be written in matricial form as:
\begin{equation}
\begin{bmatrix}
\frac{4I_{\mathrm{env}}-I_{np}}{I_{np}}\\
0
\end{bmatrix}
= 
\begin{bmatrix}
c_A & s_A \cos \varphi \\
s_A \cos \varphi & -c_A
\end{bmatrix}
\begin{bmatrix}
c_P \\
s_P
\end{bmatrix},
\end{equation}
and can be solved for $c_P$ and $s_P$. The matrix cannot be inverted if $\theta_A=\frac{\pi}{4} (\frac{\pi}{2})$ and $\varphi = \frac{\pi}{2} (\pi),$ in which case $I_{\mathrm{env}}=\frac{1}{4}I_{np}$. Otherwise the system can be inverted, yielding values for $c_P$ and $s_P$ leading to a quadratic equation by expressing: $c_P^2+s_P^2=1$. The quadratic equation has two solutions in the general case leading to upper and lower envelopes:
\begin{equation}
\label{eq:env}
  \left\{\begin{array}{@{}l@{}}
    I_{\mathrm{low}}(\theta_A,\delta\varphi) = \frac{I_{np}}{4}-\frac{I_{np}}{4}\sqrt{1-\sin^2\left( \delta \varphi \right)\sin^2\left(2\theta_A \right)}\\
    I_{\mathrm{up}}(\theta_A,\delta\varphi) \;= \frac{I_{np}}{4}+\frac{I_{np}}{4}\sqrt{1-\sin^2\left( \delta \varphi \right)\sin^2\left(2\theta_A \right)}.
  \end{array}\right.
\end{equation}
\begin{figure}[h]
\includegraphics[width=8.7cm]{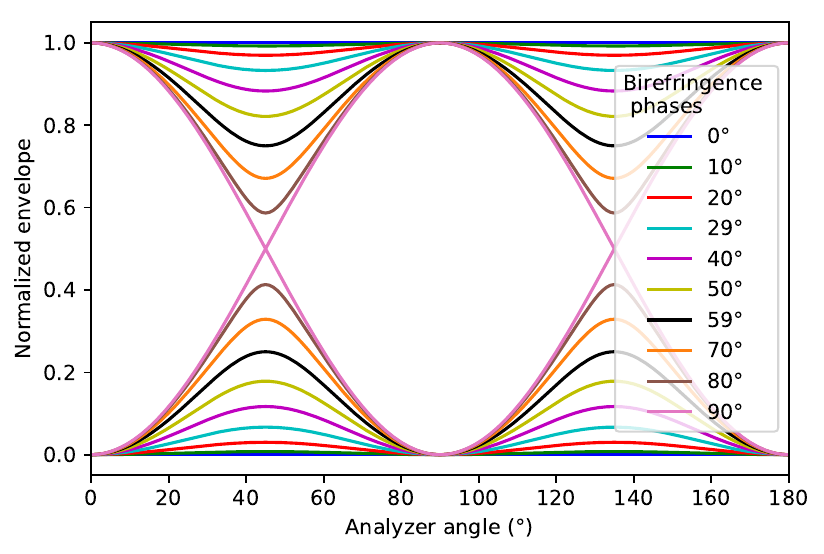}
\caption{Envelopes of the Malus laws given by Eq.~\ref{eq:env} for various birefringence differential phases.}
\label{fig:env}
\end{figure}
These expressions depend on the birefringence phase $\delta \varphi$ and provide a method to measure it. Some examples for various birefringence differential phases are plotted in Fig.~\ref{fig:env}. In the particular case of $\delta \varphi = 0(\pi)$, $ I_{\mathrm{low}}$ and $I_{\mathrm{up}}$ are constant, and are equal to 0 and 1, 
respectively. In the absence of birefringence, the generalized Malus laws systematically reach the maximum amplitude and in the case where the fiber length is a multiple of half of the beat length of birefringence, birefringence has no effect on the generalized Malus law. When birefringence is small, as in the case of the GRAVITY fibers, a modulation of the envelope of the generalized Malus laws is therefore a sign of birefringence, and can be used to measure it and predict its effects on the measurements made with GRAVITY.


\begin{acknowledgements}
GRAVITY has been developed in a collaboration by the Max Planck Institute for Extraterrestrial Physics, LESIA of Paris Observatory-PSL/CNRS/Sorbonne Universit\'e/Universit\'e Paris-Cit\'e and IPAG of Universit\'e Grenoble Alpes/CNRS, the Max Planck Institute for Astronomy, the University of Cologne, the Centro Multidisciplinar de Astrofisica Lisbon and Porto, and the European Southern Observatory. The Fiber Control Unit has been developed and procured thanks to the funding support of CNRS/INSU (CSAA and ASHRA).
\end{acknowledgements}

%
%

%
%
%
%
%
%
%
%
%

\bibliographystyle{aa} 
\bibliography{references}

\newcommand{\noop}[1]{}
\begin{thebibliography}{44}
\expandafter\ifx\csname natexlab\endcsname\relax\def\natexlab#1{#1}\fi

\bibitem[{{Anugu} {et~al.}(2020){Anugu}, {Le Bouquin}, {Monnier}, {Kraus},
  {Setterholm}, {Labdon}, {Davies}, {Lanthermann}, {Gardner}, {Ennis},
  {Johnson}, {Ten Brummelaar}, {Schaefer}, \& {Sturmann}}]{Anugu2020}
{Anugu}, N., {Le Bouquin}, J.-B., {Monnier}, J.~D., {et~al.} 2020, \aj, 160,
  158

\bibitem[{{Colavita} {et~al.}(2004){Colavita}, {Swain}, {Akeson}, {Koresko}, \&
  {Hill}}]{Colavita2004}
{Colavita}, M.~M., {Swain}, M.~R., {Akeson}, R.~L., {Koresko}, C.~D., \&
  {Hill}, R.~J. 2004, \pasp, 116, 876

\bibitem[{{Connes} \& {Reynaud}(1988)}]{Connes1988}
{Connes}, P. \& {Reynaud}, F. 1988, in European Southern Observatory Conference
  and Workshop Proceedings, Vol.~29, European Southern Observatory Conference
  and Workshop Proceedings, 1117--1129

\bibitem[{{Coud{\'e} du Foresto} {et~al.}(1995){Coud{\'e} du Foresto},
  {Perrin}, \& {Boccas}}]{Coude1995}
{Coud{\'e} du Foresto}, V., {Perrin}, G., \& {Boccas}, M. 1995, \aap, 293, 278

\bibitem[{{Coud{\'e} du Foresto} {et~al.}(1997){Coud{\'e} du Foresto},
  {Ridgway}, \& {Mariotti}}]{Foresto1997}
{Coud{\'e} du Foresto}, V., {Ridgway}, S., \& {Mariotti}, J.~M. 1997, \aaps,
  121, 379

\bibitem[{{Coud{\'e} du Foresto} \& {Ridgway}(1992)}]{Foresto1992}
{Coud{\'e} du Foresto}, V. \& {Ridgway}, S.~T. 1992, in European Southern
  Observatory Conference and Workshop Proceedings, Vol.~39, European Southern
  Observatory Conference and Workshop Proceedings, 731

\bibitem[{Faustini \& Martini(1997)}]{Faustini1997}
Faustini, L. \& Martini, G. 1997, Journal of Lightwave Technology, 15, 671

\bibitem[{{Froehly}(1981)}]{Froehly1981}
{Froehly}, C. 1981, in Scientific Importance of High Angular Resolution at
  Infrared and Optical Wavelengths, 285--293

\bibitem[{{Gardner} {et~al.}(2021){Gardner}, {Monnier}, {Fekel}, {Schaefer},
  {Johnson}, {Le Bouquin}, {Kraus}, {Anugu}, {Setterholm}, {Labdon}, {Davies},
  {Lanthermann}, {Ennis}, {Ireland}, {Kratter}, {Ten Brummelaar}, {Sturmann},
  {Sturmann}, {Farrington}, {Gies}, {Klement}, \& {Adams}}]{Gardner2021}
{Gardner}, T., {Monnier}, J.~D., {Fekel}, F.~C., {et~al.} 2021, \aj, 161, 40

\bibitem[{{Glindemann} {et~al.}(2001){Glindemann}, {Bauvir}, {van Boekel},
  {Correia}, {Delplancke}, {Derie}, {di Folco}, {Gennai}, {Gitton}, {Huxley},
  {Housen}, {Kervella}, {Koehler}, {L{\'e}v{\^e}que}, {M{\'e}nardi}, {Morel},
  {Paresce}, {Phan Duc}, {Richichi}, {Sch{\"o}ller}, {Tarenghi}, {Wallander},
  {Wilhelm}, \& {Wittkowski}}]{Glindemann2001}
{Glindemann}, A., {Bauvir}, B., {van Boekel}, R., {et~al.} 2001, in Liege
  International Astrophysical Colloquia, Vol.~36, Liege International
  Astrophysical Colloquia, ed. J.~{Surdej}, J.~P. {Swings}, D.~{Caro}, \&
  A.~{Detal}, 27--36

\bibitem[{{GRAVITY Collaboration} {et~al.}(2017){GRAVITY Collaboration},
  {Abuter}, {Accardo}, {Amorim}, {Anugu}, {{\'A}vila}, {Azouaoui}, {Benisty},
  {Berger}, {Blind}, {Bonnet}, {Bourget}, {Brandner}, {Brast}, {Buron},
  {Burtscher}, {Cassaing}, {Chapron}, {Choquet}, {Cl{\'e}net}, {Collin},
  {Coud{\'e} Du Foresto}, {de Wit}, {de Zeeuw}, {Deen},
  {Delplancke-Str{\"o}bele}, {Dembet}, {Derie}, {Dexter}, {Duvert}, {Ebert},
  {Eckart}, {Eisenhauer}, {Esselborn}, {F{\'e}dou}, {Finger}, {Garcia}, {Garcia
  Dabo}, {Garcia Lopez}, {Gendron}, {Genzel}, {Gillessen}, {Gonte}, {Gordo},
  {Grould}, {Gr{\"o}zinger}, {Guieu}, {Haguenauer}, {Hans}, {Haubois}, {Haug},
  {Haussmann}, {Henning}, {Hippler}, {Horrobin}, {Huber}, {Hubert}, {Hubin},
  {Hummel}, {Jakob}, {Janssen}, {Jochum}, {Jocou}, {Kaufer}, {Kellner},
  {Kendrew}, {Kern}, {Kervella}, {Kiekebusch}, {Klein}, {Kok}, {Kolb}, {Kulas},
  {Lacour}, {Lapeyr{\`e}re}, {Lazareff}, {Le Bouquin}, {L{\`e}na}, {Lenzen},
  {L{\'e}v{\^e}que}, {Lippa}, {Magnard}, {Mehrgan}, {Mellein}, {M{\'e}rand},
  {Moreno-Ventas}, {Moulin}, {M{\"u}ller}, {M{\"u}ller}, {Neumann}, {Oberti},
  {Ott}, {Pallanca}, {Panduro}, {Pasquini}, {Paumard}, {Percheron}, {Perraut},
  {Perrin}, {Pfl{\"u}ger}, {Pfuhl}, {Phan Duc}, {Plewa}, {Popovic}, {Rabien},
  {Ram{\'\i}rez}, {Ramos}, {Rau}, {Riquelme}, {Rohloff}, {Rousset},
  {Sanchez-Bermudez}, {Scheithauer}, {Sch{\"o}ller}, {Schuhler}, {Spyromilio},
  {Straubmeier}, {Sturm}, {Suarez}, {Tristram}, {Ventura}, {Vincent},
  {Waisberg}, {Wank}, {Weber}, {Wieprecht}, {Wiest}, {Wiezorrek}, {Wittkowski},
  {Woillez}, {Wolff}, {Yazici}, {Ziegler}, \& {Zins}}]{GRAVITY_1st_light_2017}
{GRAVITY Collaboration}, {Abuter}, R., {Accardo}, M., {et~al.} 2017, \aap, 602,
  A94

\bibitem[{{GRAVITY+ Collaboration} {et~al.}(2022){GRAVITY+ Collaboration},
  {Abuter}, {Allouche}, {Amorim}, {Bailet}, {Baub{\"o}ck}, {Berger}, {Berio},
  {Bigioli}, {Boebion}, {Bolzer}, {Bonnet}, {Bourdarot}, {Bourget}, {Brandner},
  {Cl{\'e}net}, {Courtney-Barrer}, {Dallilar}, {Davies}, {Defr{\`e}re},
  {Delboulb{\'e}}, {Delplancke}, {Dembet}, {de Zeeuw}, {Drescher}, {Eckart},
  {{\'E}douard}, {Eisenhauer}, {Fabricius}, {Feuchtgruber}, {Finger},
  {F{\"o}rster Schreiber}, {Garcia}, {Garcia}, {Gao}, {Gendron}, {Genzel},
  {Gil}, {Gillessen}, {Gomes}, {Gont{\'e}}, {Gouvret}, {Guajardo}, {Guieu},
  {Hartl}, {Haubois}, {Hau{\ss}mann}, {Hei{\ss}el}, {Henning}, {Hippler},
  {H{\"o}nig}, {Horrobin}, {Hubin}, {Jacqmart}, {Jochum}, {Jocou}, {Kaufer},
  {Kervella}, {Korhonen}, {Kreidberg}, {Lacour}, {Lagarde}, {Lai},
  {Lapeyr{\`e}re}, {Laugier}, {Le Bouquin}, {Leftley}, {L{\'e}na}, {Lutz},
  {Mang}, {Marcotto}, {Maurel}, {M{\'e}rand}, {Millour}, {More}, {Nowacki},
  {Nowak}, {Oberti}, {Ott}, {Pallanca}, {Pasquini}, {Paumard}, {Perraut},
  {Perrin}, {Petrov}, {Pfuhl}, {Pourr{\'e}}, {Rabien}, {Rau}, {Robbe-Dubois},
  {Rochat}, {Salman}, {Sch{\"o}ller}, {Schubert}, {Schuhler}, {Shangguan},
  {Shimizu}, {Scheithauer}, {Sevin}, {Soulez}, {Spang}, {Stadler}, {Stadler},
  {Straubmeier}, {Sturm}, {Tacconi}, {Tristram}, {Vincent}, {von Fellenberg},
  {Uysal}, {Widmann}, {Wieprecht}, {Wiezorrek}, {Woillez}, {Yazici}, {Young},
  \& {Zins}}]{GRAVITY_Wide_2022}
{GRAVITY+ Collaboration}, {Abuter}, R., {Allouche}, F., {et~al.} 2022, \aap,
  665, A75

\bibitem[{{GRAVITY Collaboration} {et~al.}(2023){GRAVITY Collaboration},
  {Widmann}, {Haubois}, {Schuhler}, {Pfuhl}, {Eisenhauer}, {Gillessen},
  {Aimar}, \& {et al.}}]{GRAVITY_polarization_2023}
{GRAVITY Collaboration}, {Widmann}, F., {Haubois}, X., {et~al.} 2023, \aap,
  submitted

\bibitem[{{Kervella} {et~al.}(2000){Kervella}, {Coud{\'e} du Foresto},
  {Glindemann}, \& {Hofmann}}]{Kervella2000}
{Kervella}, P., {Coud{\'e} du Foresto}, V., {Glindemann}, A., \& {Hofmann}, R.
  2000, in Society of Photo-Optical Instrumentation Engineers (SPIE) Conference
  Series, Vol. 4006, Interferometry in Optical Astronomy, ed. P.~{L{\'e}na} \&
  A.~{Quirrenbach}, 31--42

\bibitem[{{Kotani} {et~al.}(2005){Kotani}, {Perrin}, {Vergnole}, {Woillez}, \&
  {Guerin}}]{Kotani2005}
{Kotani}, T., {Perrin}, G., {Vergnole}, S., {Woillez}, J., \& {Guerin}, J.
  2005, \ao, 44, 5029

\bibitem[{{Lacour} {et~al.}(2019){Lacour}, {Dembet}, {Abuter}, {F{\'e}dou},
  {Perrin}, {Choquet}, {Pfuhl}, {Eisenhauer}, {Woillez}, {Cassaing},
  {Wieprecht}, {Ott}, {Wiezorrek}, {Tristram}, {Wolff}, {Ram{\'\i}rez},
  {Haubois}, {Perraut}, {Straubmeier}, {Brandner}, \& {Amorim}}]{Lacour2019}
{Lacour}, S., {Dembet}, R., {Abuter}, R., {et~al.} 2019, \aap, 624, A99

\bibitem[{{Lazareff} {et~al.}(2012){Lazareff}, {Le Bouquin}, \&
  {Berger}}]{Lazareff2012}
{Lazareff}, B., {Le Bouquin}, J.~B., \& {Berger}, J.~P. 2012, \aap, 543, A31

\bibitem[{{Le Bouquin} {et~al.}(2011){Le Bouquin}, {Berger}, {Lazareff},
  {Zins}, {Haguenauer}, {Jocou}, {Kern}, {Millan-Gabet}, {Traub}, {Absil},
  {Augereau}, {Benisty}, {Blind}, {Bonfils}, {Bourget}, {Delboulbe},
  {Feautrier}, {Germain}, {Gitton}, {Gillier}, {Kiekebusch}, {Kluska},
  {Knudstrup}, {Labeye}, {Lizon}, {Monin}, {Magnard}, {Malbet}, {Maurel},
  {M{\'e}nard}, {Micallef}, {Michaud}, {Montagnier}, {Morel}, {Moulin},
  {Perraut}, {Popovic}, {Rabou}, {Rochat}, {Rojas}, {Roussel}, {Roux},
  {Stadler}, {Stefl}, {Tatulli}, \& {Ventura}}]{LeBouquin2011}
{Le Bouquin}, J.~B., {Berger}, J.~P., {Lazareff}, B., {et~al.} 2011, \aap, 535,
  A67

\bibitem[{{Lefevre}(1980)}]{Lefevre1980}
{Lefevre}, H.~C. 1980, Electronics Letters, 16, 778

\bibitem[{{Lippa} {et~al.}(2018){Lippa}, {Gillessen}, {Blind}, {Kok},
  {Perraut}, {Jocou}, {Eisenhauer}, {Pfuhl}, {Haug}, {Kellner}, {Hau{\ss}mann},
  {Plattner}, {Rau}, {Hans}, {Wieprecht}, {Ott}, {Wiezorrek}, {Sturm}, {Buron},
  {Lacour}, {Genzel}, {Perrin}, {Brandner}, {Straubmeier}, \&
  {Amorim}}]{Lippa2018}
{Lippa}, M., {Gillessen}, S., {Blind}, N., {et~al.} 2018, in Society of
  Photo-Optical Instrumentation Engineers (SPIE) Conference Series, Vol. 10701,
  Optical and Infrared Interferometry and Imaging VI, ed. M.~J.
  {Creech-Eakman}, P.~G. {Tuthill}, \& A.~{M{\'e}rand}, 107011Y

\bibitem[{{Lippa} {et~al.}(2016){Lippa}, {Gillessen}, {Blind}, {Kok},
  {Yaz{\i}c{\i}}, {Weber}, {Pfuhl}, {Haug}, {Kellner}, {Wieprecht},
  {Eisenhauer}, {Genzel}, {Hans}, {Hau{\ss}mann}, {Huber}, {Kratschmann},
  {Ott}, {Plattner}, {Rau}, {Sturm}, {Waisberg}, {Wiezorrek}, {Perrin},
  {Perraut}, {Brandner}, {Straubmeier}, \& {Amorim}}]{Lippa2016}
{Lippa}, M., {Gillessen}, S., {Blind}, N., {et~al.} 2016, in Society of
  Photo-Optical Instrumentation Engineers (SPIE) Conference Series, Vol. 9907,
  Optical and Infrared Interferometry and Imaging V, ed. F.~{Malbet}, M.~J.
  {Creech-Eakman}, \& P.~G. {Tuthill}, 990722

\bibitem[{Marcuse(1976{\natexlab{a}})}]{Marcuse1976a}
Marcuse, D. 1976{\natexlab{a}}, J. Opt. Soc. Am., 66, 216

\bibitem[{Marcuse(1976{\natexlab{b}})}]{Marcuse1976b}
Marcuse, D. 1976{\natexlab{b}}, J. Opt. Soc. Am., 66, 311

\bibitem[{{Mourard} {et~al.}(2022){Mourard}, {Berio}, {Pannetier}, {Nardetto},
  {Allouche}, {Bailet}, {Dejonghe}, {Geneslay}, {Jacqmart}, {Lagarde},
  {Lecron}, {Morand}, {Rousseau}, {Salabert}, {Spang}, {Albrecht}, {Anugu},
  {Bourg{\`e}s}, {ten Brummelaar}, {Creevey}, {Deheuvels}, {Domiciano de
  Souza}, {Gies}, {Ligi}, {Mella}, {Perraut}, {Schaefer}, \&
  {Wittkowski}}]{Mourard2022}
{Mourard}, D., {Berio}, P., {Pannetier}, C., {et~al.} 2022, in Society of
  Photo-Optical Instrumentation Engineers (SPIE) Conference Series, Vol. 12183,
  Optical and Infrared Interferometry and Imaging VIII, ed. A.~{M{\'e}rand},
  S.~{Sallum}, \& J.~{Sanchez-Bermudez}, 1218308

\bibitem[{{Neumann}(1988)}]{Neumann1988}
{Neumann}, E.-G. 1988, {Single-mode fibers} (Germany: Springer Berlin,
  Heidelberg)

\bibitem[{{Peng} {et~al.}(2017){Peng}, {Cha}, {Zhang}, {Li}, \&
  {Ye}}]{Peng2017}
{Peng}, X., {Cha}, Y., {Zhang}, H., {Li}, Y., \& {Ye}, J. 2017, Optical
  Engineering, 56, 066102

\bibitem[{{Perraut} {et~al.}(2018){Perraut}, {Jocou}, {Berger}, {Chabli},
  {Cardin}, {Chamiot-Maitral}, {Delboulb{\'e}}, {Eisenhauer}, {Gamb{\'e}rini},
  {Gillessen}, {Guieu}, {Guerrero}, {Haug}, {Hausmann}, {Joulain}, {Kervella},
  {Labeye}, {Lacour}, {Lanthermann}, {Lapras}, {Le Bouquin}, {Lippa},
  {Magnard}, {Moulin}, {No{\"e}l}, {Nolot}, {Patru}, {Perrin}, {Pfuhl},
  {Pocas}, {Poulain}, {Scibetta}, {Stadler}, {Templier}, {Ventura}, {Vizioz},
  {Amorim}, {Brandner}, \& {Straubmeier}}]{Perraut2018}
{Perraut}, K., {Jocou}, L., {Berger}, J.~P., {et~al.} 2018, \aap, 614, A70

\bibitem[{{Perrin} {et~al.}(1998){Perrin}, {Coud{\'e} du Foresto}, {Ridgway},
  {Mariotti}, {Traub}, {Carleton}, \& {Lacasse}}]{Perrin1998}
{Perrin}, G., {Coud{\'e} du Foresto}, V., {Ridgway}, S.~T., {et~al.} 1998,
  \aap, 331, 619

\bibitem[{{Perrin} {et~al.}(2006){Perrin}, {Woillez}, {Lai}, {Gu{\'e}rin},
  {Kotani}, {Wizinowich}, {Le Mignant}, {Hrynevych}, {Gathright}, {L{\'e}na},
  {Chaffee}, {Vergnole}, {Delage}, {Reynaud}, {Adamson}, {Berthod}, {Brient},
  {Collin}, {Cr{\'e}tenet}, {Dauny}, {Del{\'e}glise}, {F{\'e}dou},
  {Goeltzenlichter}, {Guyon}, {Hulin}, {Marlot}, {Marteaud}, {Melse},
  {Nishikawa}, {Reess}, {Ridgway}, {Rigaut}, {Roth}, {Tokunaga}, \&
  {Ziegler}}]{Perrin2006}
{Perrin}, G., {Woillez}, J., {Lai}, O., {et~al.} 2006, Science, 311, 194

\bibitem[{Poulain {et~al.}(1975)Poulain, Poulain, \& Lucas}]{Poulain1975}
Poulain, M., Poulain, M., \& Lucas, J. 1975, Materials Research Bulletin, 10,
  243

\bibitem[{Rabien {et~al.}(2008)Rabien, Gillessen, Ziegleder, Thiel, Gr{\"a}ter,
  Haug, Eisenhauer, Perrin, Brandner, \& Straubmeier}]{Rabien2008}
Rabien, S., Gillessen, S., Ziegleder, J., {et~al.} 2008, in Optical and
  Infrared Interferometry, ed. M.~Sch{\"o}ller, W.~C. Danchi, \& F.~Delplancke,
  Vol. 7013, International Society for Optics and Photonics (SPIE), 70130I

\bibitem[{{Reynaud} {et~al.}(1992){Reynaud}, {Alleman}, \&
  {Connes}}]{Reynaud1992}
{Reynaud}, F., {Alleman}, J.~J., \& {Connes}, P. 1992, \ao, 31, 3736

\bibitem[{{Reynaud} \& {Delaire}(1993)}]{Reynaud1993b}
{Reynaud}, F. \& {Delaire}, E. 1993, Electronics Letters, 29, 1718

\bibitem[{{Rousselet-Perraut} {et~al.}(1996){Rousselet-Perraut}, {Vakili}, \&
  {Mourard}}]{Perraut1996}
{Rousselet-Perraut}, K., {Vakili}, F., \& {Mourard}, D. 1996, Optical
  Engineering, 35, 2943

\bibitem[{Setterholm {et~al.}(2023)Setterholm, Monnier, Bouquin, Anugu, Ennis,
  Jocou, Ibrahim, Kraus, Anderson, Chhabra, Codron, Farrington, Flores,
  Gardner, Gutierrez, Lanthermann, Majoinen, Mortimer, Schaefer, Scott, ten
  Brummelaar, \& Vargas}]{Setterholm2023}
Setterholm, B.~R., Monnier, J.~D., Bouquin, J.-B.~L., {et~al.} 2023, Journal of
  Astronomical Telescopes, Instruments, and Systems, 9, 025006

\bibitem[{{Shaklan} \& {Roddier}(1988)}]{Shaklan1988}
{Shaklan}, S. \& {Roddier}, F. 1988, \ao, 27, 2334

\bibitem[{{Shaklan} \& {Roddier}(1987)}]{shaklan1987}
{Shaklan}, S.~B. \& {Roddier}, F. 1987, \ao, 26, 2159

\bibitem[{{Simohamed} {et~al.}(1996){Simohamed}, {Delage}, \&
  {Reynaud}}]{Simohamed1996}
{Simohamed}, L.~M., {Delage}, L., \& {Reynaud}, F. 1996, Pure Applied Optics,
  5, 1005

\bibitem[{{Simohamed} \& {Reynaud}(1997)}]{Simohamed1997}
{Simohamed}, L.~M. \& {Reynaud}, F. 1997, Pure Applied Optics, 6, L37

\bibitem[{Tamura {et~al.}(2018)Tamura, Sakuma, Morita, Suzuki, Yamamoto,
  Shimada, Honma, Sohma, Fujii, \& Hasegawa}]{Tamura2018}
Tamura, Y., Sakuma, H., Morita, K., {et~al.} 2018, J. Lightwave Technol., 36,
  44

\bibitem[{{Vergnole} {et~al.}(2004){Vergnole}, {Delage}, \&
  {Reynaud}}]{Vergnole2004}
{Vergnole}, S., {Delage}, L., \& {Reynaud}, F. 2004, Optics Communications,
  232, 31

\bibitem[{Zhao {et~al.}(1995)Zhao, Mariotti, Foresto, Léna, \&
  Reynaud}]{Zhao1995b}
Zhao, P., Mariotti, J.-M., Foresto, V. C.~D., Léna, P., \& Reynaud, F. 1995,
  Journal of Modern Optics, 42, 2533

\bibitem[{{Zhao} {et~al.}(1995){Zhao}, {Mariotti}, {L{\'e}na}, {Coud{\'e} du
  Foresto}, \& {Maz{\'e}}}]{Zhao1995a}
{Zhao}, P., {Mariotti}, J.~M., {L{\'e}na}, P., {Coud{\'e} du Foresto}, V., \&
  {Maz{\'e}}, G. 1995, \ao, 34, 4200

\bibitem[{{Zhao} {et~al.}(1994){Zhao}, {Mariotti}, {L{\'e}na}, {Coud{\'e} du
  Foresto}, \& {Zhou}}]{Zhao1994}
{Zhao}, P., {Mariotti}, J.~M., {L{\'e}na}, P., {Coud{\'e} du Foresto}, V., \&
  {Zhou}, B. 1994, Optics Communications, 110, 497

\end{thebibliography}

\end{document}